\documentclass[10pt,journal,compsoc]{IEEEtran}

\ifCLASSOPTIONcompsoc
  \usepackage[nocompress]{cite}
\else
  \usepackage{cite}
\fi

\usepackage{amsmath}

\usepackage{graphicx}
\usepackage{times}
\usepackage{subcaption}
\usepackage{gensymb} 
\usepackage{rotating}
\usepackage{pdflscape}
\usepackage[misc]{ifsym}
\usepackage[switch]{lineno}
\usepackage{amssymb}
\usepackage{CJKutf8}
\usepackage[utf8]{inputenc}
\usepackage[ruled,vlined]{algorithm2e}

\usepackage{amsfonts}
\usepackage{colortbl}
\usepackage{adjustbox}
\usepackage{makecell}

\usepackage[dvipsnames]{xcolor}
\usepackage{hyperref}

\definecolor{Gray}{gray}{0.9}

\begin{document}
%
\title{Optimisation of non-pharmaceutical measures in COVID-19 growth via neural networks}
%
%
%
%

\author{Annalisa Riccardi$^{1}$,
        Jessica Gemignani$^{2,3}$,
        Francisco Fern\'{a}ndez-Navarro$^{4}$,~\IEEEmembership{Member,~IEEE},
        Anna Heffernan$^{\ast5,6}$,\\
        \vspace{0.5cm}
            
        \normalsize{$^{1}$Department of Mechanical and Aerospace Engineering, University of Strathclyde, Glasgow, UK}\\
\normalsize{$^{2}$Integrative Neuroscience and Cognition Center, Université de Paris, Paris, France}\\
\normalsize{$^{3}$Integrative Neuroscience and Cognition Center,CNRS, Paris, France}\\
\normalsize{$^{4}$Department of Quantitative Methods, Universidad Loyola Andaluc\'{i}a, Spain}\\
\normalsize{$^{5}$} Department of Physics, University of Guelph, Ontario, Canada\\
\normalsize{$^{6}$} Perimeter Institute of Theoretical Physics, Waterloo, Ontario, Canada \\

\IEEEcompsocitemizethanks{
\IEEEcompsocthanksitem $^\ast$To whom correspondence should be addressed; E-mail:  aheffernan@perimeterinstitute.ca
\IEEEcompsocthanksitem All authors contributed equally to this work.}
\thanks{Manuscript received July 14, 2020. \copyright 2020 IEEE.  Personal use of this material is permitted.  Permission from IEEE must be obtained for all other uses, in any current or future media, including reprinting/republishing this material for advertising or promotional purposes, creating new collective works, for resale or redistribution to servers or lists, or reuse of any copyrighted component of this work in other works.}
    
}

%
%

\markboth{Journal of \LaTeX\ Class Files,~Vol.~X, No.~Y, xxx~xxx}%
{Riccardi \MakeLowercase{\textit{et al.}}: Optimisation of non-pharmaceutical measures in COVID-19 growth via neural networks}
%



\IEEEtitleabstractindextext{%
\begin{abstract}
On $19^{\mbox{th}}$ March, the World Health Organisation declared a pandemic. Through this global spread, many nations have witnessed exponential growth of confirmed cases brought under control by severe mass quarantine or \emph{lockdown} measures. However, some have, through a different timeline of actions, prevented this exponential growth. Currently as some continue to tackle growth, others attempt to safely lift restrictions whilst avoiding a resurgence. This study seeks to quantify the impact of government actions in mitigating viral transmission of SARS-CoV-2 by a novel soft computing approach that makes concurrent use of a neural network model, to predict the daily slope increase of cumulative infected, and an optimiser, with a parametrisation of the government restriction time series, to understanding the best set of mitigating actions. Data for two territories, Italy and Taiwan, have been gathered to model government restrictions in traveling, testing and enforcement of social distance measures as well as people connectivity and adherence to government actions. It is found that a larger and earlier testing campaign with tighter entry restrictions benefit both regions, resulting in significantly less confirmed cases. Interestingly, this scenario couples with an earlier but milder implementation of nationwide restrictions for Italy, thus supporting Taiwan's lack of nationwide lockdown. The results, found with a purely data-driven approach, are in line with the main findings of mathematical epidemiological models, proving that the proposed approach has value and that the data alone contains valuable knowledge to inform decision makers.
\end{abstract}

}

\maketitle

\IEEEdisplaynontitleabstractindextext

%
\IEEEpeerreviewmaketitle
%
%
%
%
\begin{CJK}{UTF8}{gbsn}

\IEEEraisesectionheading{\section{Introduction}\label{sec:introduction}}

\IEEEPARstart{O}{n} $31^{\mbox{st}}$ December 2019, China reported to the World Health Organisation (WHO) the detection of pneumonia with an unknown etiology in the city of Wuhan \cite{10665-330760}. The viral agent was identified as a novel coronavirus, subsequently named SARS-CoV-2 due to genetic similarities to SARS-CoV, that is almost identical (96\% whole-genome level) to a known bat coronavirus \cite{LU2020565, nature1, nature2}. Recognising the growing epidemic, China banned travel to and from Wuhan and activated a national emergency response on $23^{\mbox{rd}}$ January, invoking travel and social distancing restrictions on a national scale \cite{Tian638,Chinazzi395}. Despite these actions, the virus successfully spread on a global level and by $11^{\mbox{th}}$ March, the WHO had declared a pandemic \cite{10665-331475}. The ability of this virus to spread despite acute public awareness and control actions is currently attributed, in part, to viral shedding of presymptomatic cases \cite{nature_med, Rockx1012}. With no currently approved vaccines or specific treatments, non-pharmaceutical measures are the frontline of both offense and defense in `flattening the curve' and inhibiting the infamous second wave respectively \cite{Kissler860, LEUNG20201382}.

Since the initial outbreak, researchers from the machine learning community have implemented Artificial Neural Networks (ANNs) to address mainly two types of problems: (i) virus detection through images and (ii) growth forecasting. For the former, researchers aim to develop an automatic detection system in which a neural network (typically a deep learning model) learns the main characteristics of the resulting pneumonia from historical images of infected patient \cite{apostolopoulos2020covid,wang2020covid,afshar2020covid}. In the latter, researchers typically develop a machine learning system to predict the future number of confirmed cases in a country from historical data \cite{roosa2020real,yang2020modified,petropoulos2020forecasting,fanelli2020analysis}. The forecasting model typically relies on three types of deep learning networks \cite{shone2018deep}: Fully-Connected Networks (FCNs), Recurrent Neural Networks (RNNs) and Convolutional Neural Networks (CNNs).

\begin{itemize}
    \item \textbf{Fully Connected Networks (FCNs)}: A FCN is a type of ANN where the architecture is such that all the nodes in one layer are connected to the neurons in the next layer \cite{li2018fully}. For example, the combination of FCN and interior search algorithm was proposed in \cite{rizk2020covid} for predicting the future number of infected of COVID-19 people. From a similar perspective, a FCN with three hidden layers was also proposed to predict the number of cases in Mexico  \cite{torrealba2020modeling}. Finally, the FCNs was employed (hybridized in this case with the well-known SIR model) to predict the peak of COVID-19 in Spain \cite{baltas2020monte}
    \item \textbf{Recurrent Neural Networks (RNNs)}: RNNs are the most widely used ANN architecture for addressing time series forecasting in the deep learning community. In RNNs, there is a bidirectional communication between the nodes in the different layers. Thus, the nodes in a particular layer are fed with both the outputs of previous and next layers. The architecture associated to RNNs is particularly interesting as it allows the model to memorize both short and long terms relationships among the desired output and the input variables \cite{fernandez2018time}. Regarding the implementation of RNNs for COVID-19 forecasting, in \cite{yang2020modified}, the authors trained an RNN model with data from the 2003 SARS epidemic for predicting (in a SEIR model) new infections in China. In \cite{chimmula2020time}, the authors used a Long Short Term Memory Neural Network (LSTMNN) to predict the number of new cases of COVID-19 in Canada. Similarly, LSTMs were also employed in \cite{tomar2020prediction} to forecast the number of cases in India. 
    
    \item \textbf{Convolutional Neural Networks (CNNs)}: A CNN consist of one input and output layer and several hidden layers \cite{liang2018optimization}. The hidden layers of a CNNs are based on a set of convolutional layers that convolve with multiplication or other dot  \cite{wang2019gaussian}. CNNs have been traditionally employed in image classification, in particular 2-D CNNs. However, they can also be used for times series forecasting,  1D CNNs \cite{gamboa2017deep}. In the field of COVID-19 forecasting, CNNs were proposed (as the main architecture \cite{huang2020multiple} or in combination with LSTM models \cite{dutta2020cnn})  to analyze and predict the number of confirmed cases in China \cite{huang2020multiple, dutta2020cnn}. 
\end{itemize}

Unfortunately, current research on forecasting growth fails on two issues: (i) they do not include a complete set of government decision variables in the study (maybe due to their largely qualitative nature, making it difficult to encode and preprocess with neural networks) and (ii) they do not provide guidelines on how those variables could be combined to mitigate the effect of the virus (at least partially). Motivated by this, we have proposed a soft computing approach where ANNs and optimization techniques are used concurrently to determine a different and optimal scenario of mitigation measures. The novelty of the work is dual: (i) this is the first study of this kind that, by merging data from heterogeneous sources, has created a rich dataset that is able to capture not only government actions but also the reactions and adherence to the regulations of the population; (ii) the analysis of mitigating measures was carried out through a novel optimization approach where the time series of the independent variables have been parameterised. These parameters are subsequently optimised by means of an genetic algorithm for integer programming to generate the new corresponding optimal time series. The neural network model is used to predict the slope of infected people for the new time series.

This paper is organised as follows: Sec. \ref{sec:datagathering} summarises the data gathered, sources used and quantifying techniques. Sec. \ref{sec:method} outlines the neural networks models tested, data preparation and optimization method. Results are listed in Sec. \ref{sec:results} and discussed in Sec. \ref{sec:discuss}. This study is summarised in Sec. \ref{sec:conclusion} with possible future directions outlined.


\section{Data: Two territories, two trajectories}\label{sec:datagathering}

\subsection{Italy and Taiwan}

\begin{figure*}[t]
    \centering
    \includegraphics[width=\textwidth]{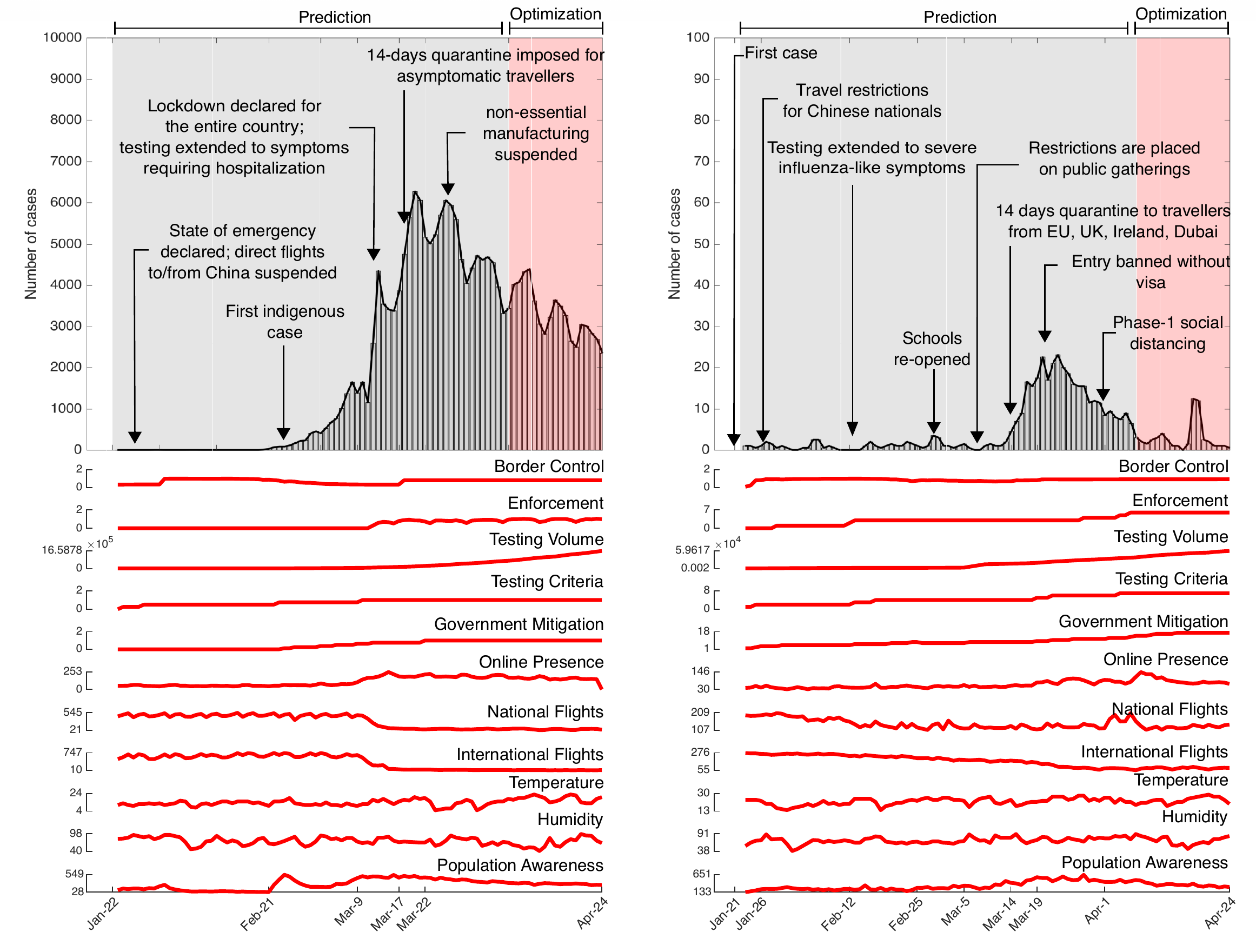}
    \caption{The top panel portrays the distributions of the growth curves in Italy (left) and Taiwan (right), along with the indication of some key moments characterizing the chronological development of the epidemics. The areas shaded in grey represent the timeframe employed by the neural networks to predict the growth, while the areas shaded in red indicate the timeframe (April 7th-24th) employed to compute the optimization of the combination of control measures via genetic algorithm. The bottom panel illustrates, for each territory, the set of variables collected and utilized to describe and predict the evolution of the epidemics.}
    \label{fig:SchemeVariables}
\end{figure*}

By mid-February, China had successfully demonstrated that the exponential growth of confirmed cases can be vastly mitigated by mass quarantine or \emph{lockdown} measures \cite{Kraemer493}. Unfortunately, by then the virus was beginning to take hold in several nations. In particular, Italy would become the next epicenter of the pandemic \cite{covid-dataset, protciv}, with dire consequences on its healthcare system; before the onset of the epidemics, Italy had approximately 5200 intensive care beds available \cite{angelico2020covid}, and efforts had to be made to increase this number during the emergency \cite{grasselli2020critical}:  on April 3rd the number of patients admitted to ICU reached an unprecedented figure of 4068 \cite{protciv}, posing a serious threat on the solidity of the entire system. In these extraordinary circumstances, the Italian government instigated severe lockdown measures similar to those implemented in China \cite{REMUZZI20201225}. 

Meanwhile Taiwan, despite its close proximity and frequent flights to China, continued to contain the growth of confirmed cases through different yet efficient and effective government measures \cite{LAI2020105946}. These actions never escalated to a nationwide lockdown. Nevertheless they successfully prevented the exponential growth predicted by unmitigated viral transmission \cite{KermackACT}, and did not overwhelm their healthcare system.

With the ultimate goal of providing a foundation that could serve policy-makers in the decision process, in regard to managing current outbreaks or preventing a future second-wave, this study aims to understand the impact and effectiveness of the possible measures a nation can employ to inhibit and constrict growth. Thus, we analyse these two different scenarios from two different regions.

\subsection{Data gathering}\label{subsec:data}
To capture and describe the evolution of the  pandemics in the two regions, several variables were employed describing the reaction of the respective governments, its timeliness, the behavior of the community, and the weather conditions. To gather the Italian data, numerous sources were used, namely the Ministry of Health \cite{trovanorme_minsalute}, the Ministry of Interior \cite{mininterno} and the Department of Civil Protection \cite{protciv}. As for Taiwan, the Taiwan Centers for Disease Control (TCDC) released timely reports to the public on government mitigation actions \cite{TaiwanCDC}. This was previously reported by \cite{10.1001/jama.2020.3151} for data available until $24^{\rm{th}}$ February 2020. We extend this data by two months until $24^{\rm{th}}$ April 2020.
Each variable was then coded to obtain a daily time series ({Fig.} \ref{fig:SchemeVariables}). The dataset is summarized below. \\

\noindent {\bf Border Control:} Several measures have been taken to control border admission.  The different levels of action applied to travellers entering the territory were equated to a numeric scale. This ranges from 0 to 5 for national citizens and 0 to 7 for non-nationals, where 0 represents zero constraints and 7 entry banned (5 equates to {14 days} quarantine). The impact of these levels is weighted by the daily cases for each territory as recorded by John Hopkins \cite{covid-dataset}. Details of the method used can be found in Supplementary Material, depicted in Tables 5-7 and {Fig.} 5.\\

\noindent {\bf Italian Enforcement:} In Italy, to supervise the compliance of the population to the restrictions, the Ministry of Interior organized a massive monitoring campaign on both individuals and businesses. Failure to comply to the restrictions could result in fines and/or criminal charges for individuals, and fines and/or suspension of the activity for businesses.
The daily report was gathered from the website of the Ministry of Interior \cite{mininterno}, and the daily number of police checks on individuals and businesses was averaged to obtain the time series \textit{Enforcement}. \\

\noindent {\bf Taiwan Enforcement:} To ensure compliance in Taiwan, Health Declaration Forms \cite{TaiwanCDChq} and mobile tracking were introduced with violators earning fines and enforced group quarantine. Table 8 in Supplementary Material lists and quantifies these consequences as a step function. \\

\noindent {\bf Testing Volume:} Starting from February $24^{\mbox{th}}$, the Italian Department of Civil Protection released a daily public report of the the number of tests administered, which was gathered from its website  \cite{protciv} and employed as a variable \emph{Testing Volume}. In Taiwan, the TCDC initiated testing in early January, by $8^{\mbox{th}}$ March, 13,855 tests had been carried out, with approximately 1000 tests a day since. Testing numbers were released in the TCDC public reports \cite{TaiwanCDC}. Days where exact numbers were not released, the daily count has been interpolated  by averaging the increase of testing volume over the `missing' days. \\

\noindent {\bf Testing Criteria:} In Italy, at the beginning of the pandemic diagnostic tests were only performed on symptomatic individuals traveling back from China. Gradually, testing has been extended to a larger part of the population. Each extension of the testing inclusion criteria was modeled as a step in the corresponding variable. The data was gathered from the website of the Ministry of Health \cite{minsalute} and presented in Table 9 in Supplementary Material.  In Taiwan, surveillance information was publicly made available by the TCDC \cite{TaiwanCDC} and previously noted in \cite{10.1001/jama.2020.3151}. Initial suspect cases were screened for 26 viruses and targeted travellers from Wuhan. As the virus spread, this was extended to symptomatic individuals with close contact to infected individuals, clusters of fever and pneumonia cases unresponsive to antibiotics. Testing of symptomatic individuals was eventually extended to all inbound travellers. The representative step function is described in Table 10 of Supplementary Material. \\

\noindent {\bf Government Mitigation:} Over the course of the emergency, the Italian Presidency of the Council of Ministers, the Ministry of Health and the Department of Civil Protection issued several decrees and ordinances of increasing severity, aimed at containing the spread of the epidemic. The first measure was the creation of several red zones in Northern Italy, while the last step was the complete nationwide lockdown, along with the suspensions of all productive businesses not directly involved in supplying essential services. Based on the succession of these measures, we modeled the \textit{Government Mitigation} variable as a step function, with each step corresponding to one of the legal acts. The end value represents the complete national lockdown. In Taiwan, general public communication including heightened hand and respiratory hygiene recommendations started on $6^{\mbox{th}}$ January through the TCDC \cite{TaiwanCDC}. Through chance, school holidays commenced on $20^{\mbox{th}}$ January, however they were extended for two weeks in response to the increased spread of the virus. From there the government mitigation actions involved heightened hygiene practices with increased social distances guidance, e.g. large gatherings, wearing masks in crowded public places, guidance for educational institutions, etc. Yet, unlike Italy, they did not go into national lockdown. Actions taken by Italy and Taiwan are listed and quantified in Tables 11 and 12 of Supplementary Material respectively. \\

\noindent {\bf Online Presence:} 
A daily measure of home confinement is equated to search query data via the publicly available service \textit{Google Trends}, using the search terms: Netflix, Amazon Prime Video, Zoom Video Communications, Skype. Corresponding search volume indices were combined into a single variable, \emph{Online Presence}. In Italy, as {Fig.} \ref{fig:SchemeVariables} illustrates, an initial peak coincides with the national lockdown (March 8th, Level 5 in Table 11), and remains at high levels until the end of our data collection ($24^{\rm{th}}$ April 2020). In Taiwan, despite no national lockdown activated, an increased home presence can be seen in mid-March in {Fig.} \ref{fig:SchemeVariables}. This coincides with an increase growth curve and hence could be interpreted as the public's personal response to the increasing number of cases in Taiwan. \\

\noindent {\bf National and International Flights:} To quantify mobility within and outwith both Italy and Taiwan, data from national and international arriving and outgoing flights have been collected from the online platform \textit{Flightradar24} \cite{flightradar24}. \\

\noindent {\bf Temperature and Humidity:} A recent study suggests weather parameters such as temperature and humidity can affect the spreading of the virus \cite{weather_ref}. Data collected from \textit{World Weather Online} \cite{weather} of 24-hours average temperature ($C^o$) and humidity ($\%$) values have been added to the data set. The online service has been queried using Taiwan and Italy longitude and latitude representative coordinates (23.7 and 121 for Taiwan, 43 and 12 for Italy). The value of humidity and temperature in a specific location in each region has been taken as representative of that territory in the period observed. \\

\noindent {\bf Population Awareness:} To quantify people awareness of the phenomena, \textit{Google Trends} can be used \cite{HUSNAYAIN2020221}. Data has been collected here by querying people searches related to the list of keywords: "coronavirus", "covid", "quarantena", "wuhan", "virus" (in Italy) and "coronavirus", "covid", "quarantine", "wuhan", "virus" (in Taiwan). For each word the top 5 related queries are collected and the results of all searches aggregated. In Italy, a peak can be observed in the following days of the first case, coinciding with the creation of red zones in Northern Italy ({Fig.} \ref{fig:SchemeVariables}).  This diminishes only to return with the national lockdown (March 8th, Level 5 in Table 11). This online presence slowly decreases until the end of our data collection ($24^{\rm{th}}$ April 2020). In Taiwan, interest also initially peaks with the first case, diminishing to peak again as growth begins to surge along with government mitigation. The full list of words used in the google trends analysis for both territories is available in Table 13 of Supplementary Material.


\section{Simulating and Optimising Growth Curves}\label{sec:method}

\subsection{Neural Network Models}
The data collected are used as independent variables for the multivariate time series analysis where the slope is the dependent variable. Three neural network models are tested: 2D Convolutional Neural Network (CNN) \cite{Goodfellow-et-al-2016}, Extreme Learning Machine (ELM) \cite{Huang2006} and Bidirectional Long Short-Term Memory Neural Network (BiLSTMNN) \cite{hochreiter1997long}. \\


\noindent \textbf{{CNN:}} 
The deep CNN proposed in this study is composed of four sliding convolutional filters with kernels equal to 8, 16, 32 and 32, each one connected in between by a batch normalisation layer to speed up training and reduce the sensitivity, and a Relu layer to set to zero any value less than zero. The output of the convolutional layers feeds into a dropout layer for regularisation, and a fully connected layer for the continuous prediction of the output, the slope. The model depends on two network parameters: the size of the CNN filter and the dropout layer probability; and three training parameters: the mini-batch size, the learning rate drop factor and drop period.\\


\noindent \textbf{ELM:}
The ELM framework was originally presented as a single hidden layer feedforward neural network, whose connections between input and hidden layers are randomly assigned. The only parameters needed to be tuned are the weights between the hidden and the output layers, which are analytically estimated by solving the standard least-squares minimization problem. This methodology significantly improves the computational burden of the algorithm and allows it to process heavy datasets in a reasonable computational time \cite{Kasun2013}. ELM models have reported competitive performances in both standard machine learning datasets (regression and classification) \cite{Huang2012} and other more challenging problems such as image detection \cite{mohammed2011human}, time series \cite{Tian2010}, topological information \cite{da2018geographical} or real-time river flow prediction \cite{yaseen2019enhanced}. In this experimental study, the ELM neural network version of the framework with sigmoidal and linear basis functions in the hidden and output layers was considered. The model critically depends on two hyper-parameters: the number of hidden neurons, $S$, and the regularization penalty term, $C$. \\


\noindent \textbf{LSTMNN:} 
Unlike the two previously described models, the LSTMNN method belongs to the family of Recurrent Neural Networks (RNNs). In traditional Feedforward Neural Networks (FFNNs), information moves only in a direct way from the input layer to the output one. RNNs are based on an architecture in which the data through the system moves constituting a direct cycle \cite{ming2017understanding}. The main difference between standard RNNs and LSTMNNs models is that the latter are capable of memorizing a time series value for an arbitrary length of time. The model used in this study is a Bidirectional LSTM (BiLSTM) network. This is usually a superior prediction approach in time series forecasting than the simple LSTM model, as it learns from the complete time series data (past and future states). The model depends on three network hyper-parameters: the number of hidden neurons in the bidirectional LSTM layer, the size of the first fully connected  layer, the dropout probability rate; and three training hyper-parameters (initial learning rate, learning rate drop period and factor).

The optimal set of hyperparameters for the neural networks have been obtained by minimising the root mean square error on the predictions by means of a Bayesian optimisation process. Bayesian optimisation is an algorithm used in global optimisation to minimise a certain objective function, treated as a black box, by varying the value of its independent variables. The algorithm itself relies on an internal Gaussian process that approximates the objective function, and is trained by subsequent evaluations of the true objective. The approximated model is used for optimisation to reduce computational costs and for its robust nature with stochastic noise in function evaluations. 

\subsection{Data preparation}
The collected data has been divided for training and testing, and reshaped as feature input for the multivariate time series analysis. Furthermore, in order to be able to use the neural network models for the subsequent analysis, the optimisation of government restrictions, the dependence of the historical values of the dependent variables is omitted
from the input features space. Hence for the ELM and BiLSTMNN models the input features are modelled as the vector 
\begin{equation}
    \mathcal{X}_t=[X(t), ..., X(t+N)]\in\mathbb{R}^{11 N}
\end{equation}
where $N$ is the time window parameters, and $X(t)$ is the vector of independent variables at time $t$ as described in the Data Collection section. Conversely for the CNN model the input space is defined as
\begin{equation}\mathcal{X}_t=\begin{bmatrix}
x_1(t) & .. & x_1(t+N) \\
... & .. & ... \\
x_{11}(t) & .. & x_{11}(t+N)
\end{bmatrix}\in\mathbb{R}^{11\times N}\end{equation}
where $x_i(t)$ is the $i$-th independent variable at time $t$.
The hyperparameter $N$ is optimised, by means of a Bayesian optimisation process, together with the networks hyperparameters.

\subsection{Optimisation}\label{section:opt}
The best trained neural network model (BiLSTMNN) is used to have a deeper understanding of government regulations and their impact on the slope function. Taking
\begin{equation}
    y_t=\mathcal{N}(\mathcal{X}_t), \quad \mathcal{X}_t\in\mathbb{R}^{M}
\end{equation}
as the neural network model, where $M = 11N$ or $M=11\times N$ depending on the model used, the following optimisation problem has been formulated
\begin{equation}
   \min J(\mathbf{x}), \qquad \mathbf{x}\in \Omega,
\end{equation}
where $J(\mathbf{x})$ is the objective function defined as the maximum value of the slope along the whole time frame considered. {
\begin{equation}
    J(\mathbf{x}) = \max_t \mathcal{N}(\mathcal{X}_{t}(\mathbf{x})).
\end{equation}
The daily slope value is computed as the angular coefficient of the linear regression of the values of daily cumulative infected of the current day and two days ahead, as a smoother representation of the daily number of infected.
}
$\mathbf{x}$ is a set of optimisation variables designed to model government restrictions previously discussed (border control, enforcement, testing volume, testing criteria, social distance, flight volume and people awareness), excluding weather parameters as they are not controllable by government. The complete set of optimisation variables for each territory is described in Table \ref{tab:Optvars}. $\Omega$ is the feasible region defined by the lower and upper bounds of the set of optimisation variables with some constraints. Inequality constraints are introduced to enforce the temporal order of the variables modelling quarantine incoming and the step functions of social distance, quarantine control (only for Taiwan) and testing criteria. To avoid the optimisation converging towards a temporal delay of the first case, an equality constraint is introduced to ensure the day of first confirmed case matches reality. The optimisation problem aims to minimize the maximum value of the slope - this can be interpreted as minimising the number of daily infections so to not put the health care system under extreme strain.

\begin{table*}[!htb]
    \centering
    {\caption{Government restriction optimisation variables list and descriptions}   \label{tab:Optvars}}
    \begin{tabular}{|| c | c | p{10cm}  ||}
        \hline \hline
Italy & Taiwan &\\
\hline
$x_1,x_2,x_3$ & $x_1,..,x_{12}$ & \textit{Quarantine incoming}: variables are dates of actions as highlighted in Tables 6 and 7 in Supplementary material.\\
$x_4,x_5$ & $x_{13}; x_{14}, ...x_{19}$& \textit{Quarantine control}: for Italy the first variable models the time and the second variable models the shift; for Taiwan the first variable models the time of the initial step and the rest of the variables model the step function interval sizes\\
$x_6$& $x_{20}$ & \textit{Testing}: the variable models the shift in time of the day in which testing had a steep increase\\
$x_7; x_8...,x_{11}$& $x_{21}; x_{22},...x_{28}$ & \textit{Testing criteria}: the first variable models the time of the initial step and the rest of the variables model the step function interval sizes\\
$x_{12}; x_{13},..,x_{19}$& $x_{29}; x_{30},..x_{46}$& \textit{Social distance}: the first variable models the time of the initial step and the rest of the variables model the step function interval size\\
$x_{20},x_{21}$& $x_{47},x_{48}$& \textit{Social distance online platform}: the first variable models the time and the second variable models the shift\\
$x_{22},x_{23}$& $x_{49},x_{50}$&\textit{Flights national}: the first variable models the time and the second variable models the shift\\
$x_{24},x_{25}$& $x_{51},x_{52}$& \textit{Flights international:} the first variable models the time and the second variable models the shift\\
$x_{26},x_{27}$& $x_{53},x_{54}$& \textit{People awareness}: the first variable models the time and the second variable models the shift\\
        \hline \hline    
        \end{tabular}
\end{table*}

For Italy the problem can be formulated mathematically as
\vspace{-0.1em}
\begin{eqnarray}
\min_\mathbf{x}{ \max_t \mathcal{N}(\mathcal{X}_t(\mathbf{x}))}\\
\textrm{s.t}\nonumber\\
A\begin{bmatrix}x_1\\x_2\\x_3\end{bmatrix}
\leq0\\
x_7+\sum_{i=8}^{11}x_i\leq T\\
x_{12}+\sum_{i=13}^{19}x_i\leq T\\
t=1, \, t=\{y_t>0 | y_t=\mathcal{N}(\mathcal{X}_t(\mathbf{x})), \forall t \}\\
\mathbf{x}\in\mathbb{N}^{27}, L\leq\mathbf{x}\leq U
\end{eqnarray}
where \begin{equation}
    A=\begin{bmatrix}
1&-1  &0 \\
0 & 1 & -1
\end{bmatrix}.
\end{equation}
is the matrix defining the linear inequality temporal constraints on the quarantine incoming variables. $T$ is the time horizon parameter, and $L,U$ are the vectors defining the lower and upper bounds of the optimisation variables. For Taiwan the problem is defined as,
\begin{eqnarray}
\min_\mathbf{x}{ \max_t \mathcal{N}(\mathcal{X}_t(\mathbf{x}))}\\
\textrm{s.t}\nonumber\\
    B_1 \begin{bmatrix}x_1\\...\\x_{..}\end{bmatrix}\leq0,\qquad
    B_2 \begin{bmatrix}x_6\\...\\x_{12}\end{bmatrix}\leq0,\\
    x_{13}+\sum_{i=14}^{19}x_i\leq T, \\
    x_{21}+\sum_{i=22}^{28}x_i\leq T, \\
    x_{29}+\sum_{i=30}^{46}x_i\leq T\\
    t=1, \, t=\{y_t>0 | y_t=\mathcal{N}(\mathcal{X}_t(\mathbf{x})), \forall t \}, \\
    \mathbf{x}\in\mathbb{N}^{54}, L\leq\mathbf{x}\leq U,
\end{eqnarray}
with 
\begin{equation}
    B_1 = \left[\begin{array}{cccccccccccc}
1&-1&0&0&0&0&0&0&0&0&0&0\\
0&0&1&-1&0&0&0&0&0&0&0&0\\
0&0&0&0&1&0&-1&0&0&0&0&0\\
0&0&0&0&0&0&0&1&-1&0&0&0\\
0&0&0&0&0&0&0&0&0&1&-1&0\\
0&0&0&0&0&0&0&0&0&0&-1&1\\
\end{array}\right],
\end{equation}
\begin{equation}
    B_2 = \begin{bmatrix}
-1&1&0&0&0&0&0\\
-1&0&1&0&0&0&0\\
-1&0&0&1&0&0&0\\
-1&0&0&0&1&0&0\\
-1&0&0&0&0&1&0\\
-1&0&0&0&0&0&1\\
\end{bmatrix},
\end{equation}
where $B_1$ and $B_2$ are the matrices defining the linear inequality temporal constraints on the quarantine incoming variables.

The integer programming Genetic Algorithm is used to solve the optimisation problem. Genetic Algorithms are stochastic global optimisation strategies that mimic the behaviour of natural biological evolution of mutation and crossover. They are initialised with a pool of potential solutions and by evolutionary operators and applying the principle of survival of the fittest, increasingly better offspring populations are generated, which terminate when no further improvements can be made. Genetic algorithms have the main advantage that they can solve problems with integer variables and the objective function is treated as a black box.

\subsection{Summary of the computational approach}

We propose a soft computing algorithm which combines neural networks and genetic algorithms, to determine (from a purely data-driven approach), the best set of mitigating actions. The algorithm is based on two independent stages that are summarized below and outlined in Algorithm. \ref{algo:softcomputingapproach}.

In the first stage, a set of neural networks models are trained to estimate the daily slope increase of cumulative infected from variables such as government actions and the reactions and adherence of the population to such regulations. The best performing model is selected taking into account the reported root mean square percentages of the neural networks implemented. The choice of training a neural network model per territory is driven by the need of capturing region specific static parameters. Time-varying parameters that have an influence on the predicted growth curve, and are not among those described in Sec.~\ref{subsec:data}, can still be captured by the network model as hidden features if they are correlated to the input variables selected.

In the second stage, the best performing neural network (BiLSTMNN) is used as a surrogate model in an optimization procedure that targets the best set of mitigating actions. The optimization procedure incorporates an ad-hoc parametrization of the time series governmental restrictions. These parameters are then optimised through genetic algorithm for integer programming to generate the new corresponding optimal time series. The algorithmic flow of this soft computing approach is provided below.

{
\begin{algorithm}
\SetAlgoLined
\KwResult{Optimised governmental measures for region under study}
 Data gathering\;
 Train BiLSTMNN model for region under study\;
 Initialise Table \ref{tab:Optvars} variables bounds for region under study\;
 Initialise GA population ($pop_0$, 100 individuals)\;
 i=0\;
 \While{$ngen < 1000$}{
   \For{each individual $\mathbf{x}$ in $pop_i$}{
   construct time series $\mathcal{X}(\mathbf{x})$\;
   evaluate predictions $\mathcal{N(\mathcal{X}(\mathbf{x}))}$\;
   compute fitness $J(\mathbf{x}) =\max_{t} \mathcal{N}(\mathcal{X}_t(\mathbf{x}))$\;
   compute inequality constraints $c(\mathbf{x})$ as in Section \ref{section:opt}\;
   }
   apply GA operators to obtain $pop_{i+1}$\;
   $ngen=ngen+1$\;
 }
 return best individual, $x_{best}$, of $pop_i$\;
 construct time series $\mathcal{X}(x_{best})$\;
 \caption{Optimisation of non-pharmaceutical measures in COVID-19 algorithm for region under study}\label{algo:softcomputingapproach}
\end{algorithm}
}


\section{Results}\label{sec:results}
\subsection{Neural Network model}
In this section, the performance of the different neural networks models for Italy and Taiwan are statistically compared. For model selection, following the recommendations reported in \cite{bergmeir2012use}, the time series were split in two parts: (i) the training set is roughly the first 2 and a half months of data; and (ii) the generalization set is the last section of each time series. The training set was used to estimate the parameters of the models and the generalization set to assess the performance of the different neural networks in unseen data \cite{bergmeir2012use}. In the experiments, the generalization set included the last five points of each time series, as suggested in \cite{huang2019multiple}. Table \ref{tab:statistics} shows the Root Mean Square Error Percentage (RMSEP) in the generalization set of the best run per model (out of 1000 runs), $\mathrm{Best}_{RMSEP}$, and the variation coefficient of the RMSEP in the generalization set for the different models implemented, $\mathrm{CV}_\mathrm{RMSEP}$. Furthermore, taking into account the set of RMSEPs of the different models over the different runs, the mean rankings of RMSEP, $\overline{R}_\mathrm{RMSEP}$, for the different neural networks models are obtained (Table \ref{tab:statistics}). From the analysis of the results, it can be concluded, from a purely descriptive point of view, that the BiLSTMNN method obtained the best results in the two problems considered in mean rankings. Additionally, the best BiLSTMNN achieved the best RMSEP in Taiwan and second best in Italy (more consistent performance, more homogeneous results, better CV).

\begin{table}[!htb]
\centering
{\caption{Statistical results of the neural networks implemented}\label{tab:statistics}}
\begin{tabular}{||cccccc||}
\hline
\hline
& \multicolumn{5}{c||}{Italy}\\ \hline
Models & $\mathrm{Best}_{RMSEP}$ & $\mathrm{CV}_\mathrm{RMSEP}$ &  & $\overline{R}_\mathrm{RMSEP}$ & p-value\\ \cline{2-6}
CNN & 0.0277  & 0.4858 &  & 6.0589 & $0.0000^{\dag}$\\         
ELM &  0.0529 & 0.3031 &  & 6.1222 & $0.0000^{\dag}$ \\
BiLSTMNN & 0.0492 & 0.3005 &  & 2.8189 & -\\
\hline
& \multicolumn{5}{c||}{Taiwan}\\ \hline
Models & $\mathrm{Best}_{RMSEP}$ & $\mathrm{CV}_\mathrm{RMSEP}$ &  & $\overline{R}_\mathrm{RMSEP}$ & p-value\\ \cline{2-6}
CNN & 0.1775 & 0.5056 &  & 4.6589  & 0.3942\\         
ELM & 0.1367 & 0.3348 &   & 6.0011 & $0.0000^{\dag}$ \\
BiLSTMNN & 0.1402 & 0.4350 &  & 4.3400 & -\\
\hline
\hline
\end{tabular}
\end{table}

In this study, hypothesis testing was used to provide statistical support for the discussion of the results. A performance analysis through parametric tests could lead to mistaken conclusions in this research study. A previous evaluation of the RMSEP values provided by the implemented methods resulted in rejecting the normality and equality of the variance hypothesis. For these reasons, nonparametric tests were implemented to determine the statistical significance of the results previously reported \cite{siegel1957nonparametric}. Specifically,  two non-parametric Friedman tests were carried out with the rankings of RMSEP of the models. For the two multivariate time series considered, the $p$-values associated to the Friedman test were smaller than 0.05 ($\alpha = 0.05$), and therefore, the null hypotheses stating that all algorithms perform equally in mean RMSEP rankings were rejected for both problems.

Based on this rejection, the nonparametric Bonferroni Dunn-Sidak test was implemented to compare all neural network methods to the BiLSTMNN method (which was used as the control method) \cite{vsidak1967rectangular}. Table \ref{tab:statistics} shows the p-value results of the Bonferroni Dunn-Sidak test for the two problems considered. Thus, the Bonferroni Dunn-Sidak's tests indicate that the control method (BiLSTMNN) statistically outperforms all the remaining models in Italy and also outperforms, in statistical terms, the ELM model for Taiwain. Furthermore, the BiLSTMNN method achieves the best mean ranking in the two problems considered, which, in our opinion, justifies our decision to use it as the base model for the next stage (optimization study). An analogous study was also performed to determine the model to be analyzed within all the different BiLSTMNN configurations.

\begin{center}
    \begin{figure}
        \centering
        \includegraphics[width=0.5\textwidth]{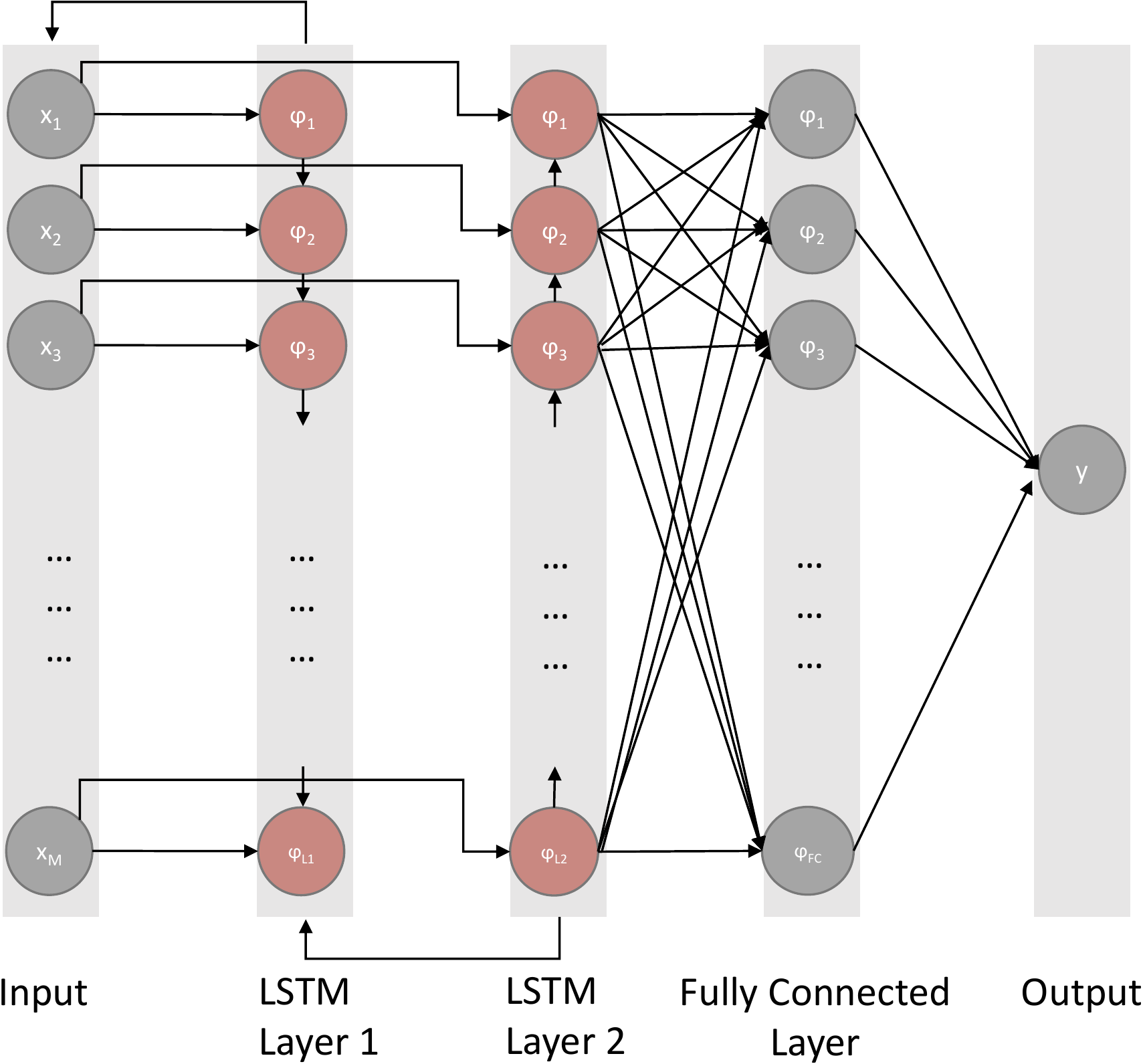}
        \caption{Graphical representation of the BiLSTM network model used for Italy and Taiwan. Parameter \textit{Number of Hidden Units} refers to the size of the LSTM layers, parameter \textit{First Layer Size} refers to the size of the fully connected layer}
        \label{fig:NNvisual}
    \end{figure}
\end{center}

The best BiLSTMNN model for each region is obtained for the optimal value of hyperparameters reported in Table \ref{tab:LSTMhyperparam}, the network visualisation model in Fig. \ref{fig:NNvisual} and the prediction model in {Fig.} \ref{fig:NNmodels}. 
\begin{table}[!htb]
    \centering
    {\caption{BiLSTMNN optimal hyperparameters 
}\label{tab:LSTMhyperparam}}
    \begin{tabular}{|| l  || c | c ||}
        \hline \hline
        BiLSTMNN & Italy & Taiwan\\
        \hline 
            Time Window & 14 & 21\\  
            Number of Hidden Units & 155& 770\\
                 First Layer Size & 469 & 107\\
     Dropout Layer probability & 0.18& 0.69\\
    Initial Learn Rate & 2.4e-04& 5.2e-06\\
    Learn Rate Drop Factor  & 0.02&  0.01 \\
      Learn Rate Drop Period & 111 & 88\\
        \hline \hline    
        \end{tabular}
\end{table}
The multivariate time series model captures the trend of the slope in both cases, with a more accurate overall error in the case of Italy (due to the less noisy nature of the Italian time series). It is important to clarify, as mentioned previously, that the lagged values of the slope, cumulative cases or daily cases were not included in the model to reduce dependence between input variables and assist the interpretation of the optimum values. These variables are usually the most discriminant in times series analysis, which highlights even more the competitive results yielded. Thus, the BiLSTMNN is the appealing approach for this problem.

\begin{center}
    \begin{figure}
        \centering
        \begin{subfigure}{0.99\columnwidth}
  \centering
  \includegraphics[width=\textwidth]{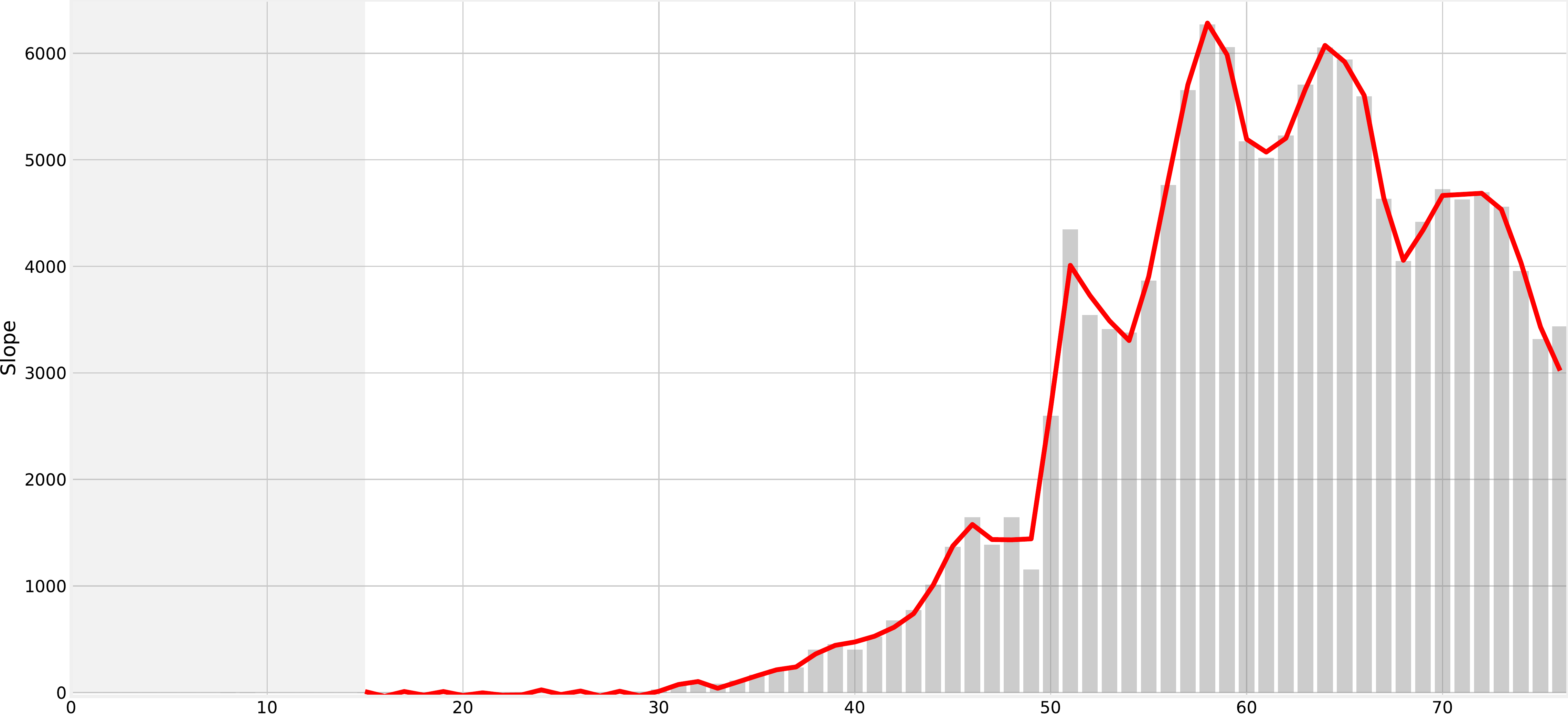}
  \caption{Italy}
  \label{fig:sub1}
\end{subfigure}
\begin{subfigure}{0.99\columnwidth}
  \centering
  \includegraphics[width=\textwidth]{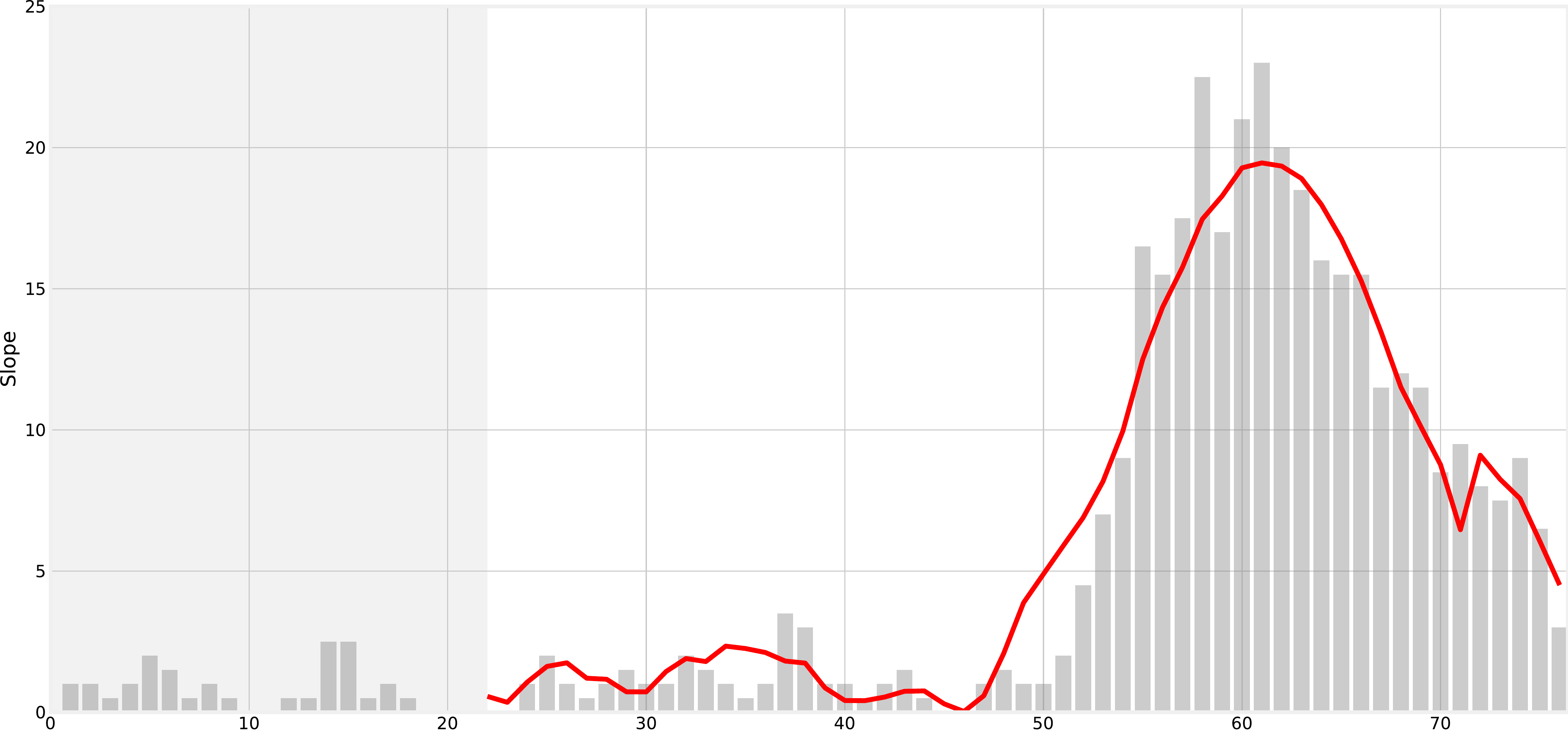}
  \caption{Taiwan}
  \label{fig:sub2}
\end{subfigure}
        \caption{Slope prediction (red) as computed by the best BiLSTMNN Neural Network model. Grey shaded area indicates model time window for which predictions cannot be made, histogram represents real slope values.}
        \label{fig:NNmodels}
    \end{figure}{}
\end{center}

\subsection{Optimisation}

\begin{center}
    \begin{figure*}
        \centering
        \includegraphics[width=\textwidth]{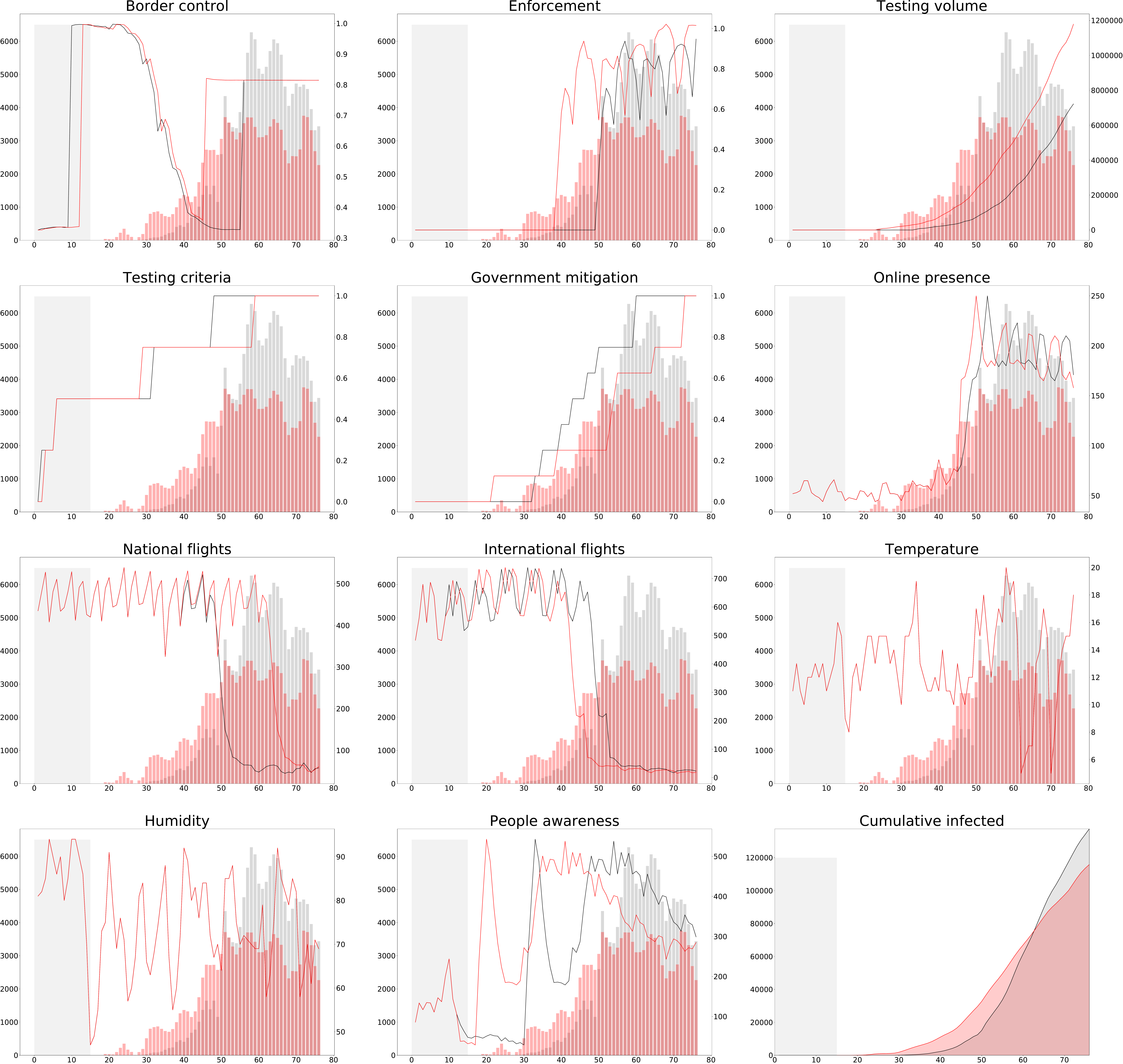}
        \caption{Optimal solution for Italy. On the right axis, black solid line represents the original variable, red solid line is the optimised solution. On the left axis the histograms of slope values for original solution (grey) and optimal one (red) are reported. Last plot represents the cumulative value of infected people in the country for original solution.}
        \label{fig:optIT}
    \end{figure*}
\end{center}

\begin{center}
    \begin{figure*}
        \centering
        \includegraphics[width=\textwidth]{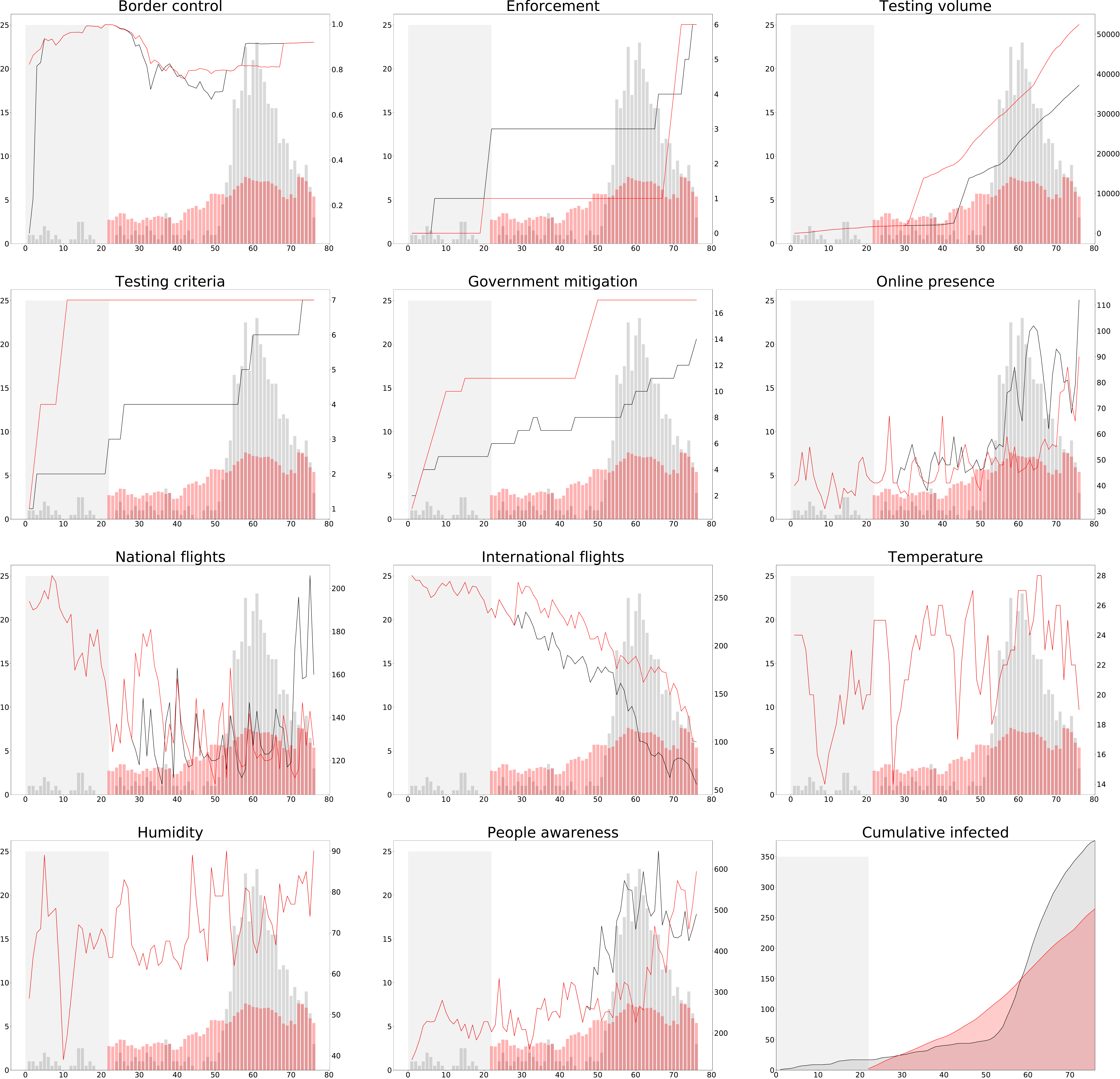}
        \caption{Optimal solution for Taiwan. On the right axis, black solid line represents the original variable, red solid line is the optimised solution. On the left axis the histograms of slope values for original solution (grey) and optimal one (red) are reported. Last plot represents the cumulative value of infected people in the region for original solution.}
        \label{fig:optTW}
    \end{figure*}
\end{center}

A genetic algorithm has been used to solve the corresponding optimisation problem for each region and promising results have been found that do not contradict previous findings of epidemiological models (Figures \ref{fig:optIT} and \ref{fig:optTW}). All variables are considered integer and optimised within the bounds reported in Table \ref{tab:OptvarsBounds}, together with their optimal value. The genetic algorithm, with population size of 100 and the maximum number of generations set to 1000, has been run with the stopping criteria set as the maximum number of generations reached or the average relative change in the best fitness function value over 100 generations being less than or equal to a tolerance of $1e-10$. 

\begin{table*}[!htb]
    \centering
    {\caption{Government restriction optimisation variables bounds and optimal values} \label{tab:OptvarsBounds}}
    \begin{tabular}{|| c | p{3cm} | p{2cm} || c | p{3cm} | p{2cm}||}
        \hline \hline
Italy & bounds & optimal value& Taiwan & bounds & optimal value\\
\hline
$x_1,x_2,x_3$ &[1,20], [5,25], [20,76]& 13, 21, 46 & $x_1,..,x_{12}$ & [1 3], [1 5], [1 5], [1 9], [24 44], [48 68], [31 51], [25 45], [27 47], [36 56], [43 63], [40 60] & 1,	1,	1,	1,	24,	68,	31,	25,	47,	36,	43,	40\\
$x_4,x_5$ &[15, 50], [-14,14]& 44, -11&$x_{13}; x_{14}, ...x_{19}$& [1,20], $[1,50]^6$& 20,	48,	1,	1,	1,	1,	1 \\
$x_6$& [20,50]& 24&$x_{20}$ & [30,60]& 31\\
$x_7; x_8...,x_{11}$&[1,10], $[1,30]^4$& 3,	3,	23,	30,	15& $x_{21}; x_{22},...x_{28}$ & [1,20], $[1,50]^7$&  1,	1,	1,	1,	5,	1,	1,	27\\
$x_{12}; x_{13},..,x_{19}$&[20,40], $[1,30]^7$ & 22,	17,	14,	1,	1,	10,	8,	3 & $x_{29}; x_{30},..x_{46}$& [1,20], $[1,30]^{17}$&1,	1,	1,	1,	1,	1,	1,	1,	1,	1,	5,	30,	1,	1,	1,	1,	1,	1\\
$x_{20},x_{21}$& [15,50], [-14,14]& 49,	-3			& $x_{47},x_{48}$& [15,70], [-14,14]& 15,	14		\\
$x_{22},x_{23}$&[15,50], [-14,14]& 26,	14&$x_{49},x_{50}$&[15,70], [-14,14]&15,	14\\
$x_{24},x_{25}$&[15,50], [-14,14]& 16,	-6&$x_{51},x_{52}$& [15,70], [-14,14]&15,	14	\\
$x_{26},x_{27}$&[15,50], [-14,14]& 26,	-13&$x_{53},x_{54}$&[15,70], [-14,14]&34,	14 \\
        \hline \hline    
        \end{tabular}
\end{table*}

The optimisation process for Italy terminates after almost 600 iterations when the change in the penalty fitness value is less than the allowable tolerance. All constraints are satisfied and the final objective function is 3755, with respect to the original value of 6272. This is almost a $40\%$ reduction with respect to the original value and equates to a reduction of almost 21,700 infected individuals at the end of the prediction horizon, corresponding to $16\%$ of the actual total value.

The optimisation process for Taiwan terminates after almost 300 iterations again when the change in the penalty fitness value is less than the allowable tolerance. All constraints are satisfied and the final objective function is 7.6, with respect to the original value of 23. This is almost a $63\%$ reduction with respect to the original vale and a reduction of 102 infected individuals at the end of the prediction horizon, corresponding to $28\%$ of the real value.
\section{Discussion} \label{sec:discuss}
\subsection{Italy}
As illustrated in {Fig.} \ref{fig:optIT}, an earlier and more extensive testing campaign (+56$\%$) could have been combined with an earlier creation of red zones in Northern Italy and, interestingly, with a milder implementation of the government mitigation plan on the rest of the country:  the localized restrictions could have been maintained as sufficient measures until $14^{\mbox{th}}$ March. Furthermore, in this scenario the country-wide suspension of manufacturing businesses, a measure burdened with a substantial socio-economic impact, is delayed and limited to a shorter period of time, while the volume of police checks, aimed at monitoring the population compliance, is larger and commences earlier. This is all combined with an earlier and more pervasive consciousness of the public about the risk posed by the epidemics and thus on the benefits of a temporary reduction of social contacts.

The containment of the growth is also optimally managed in Italy by earlier restrictive regulations on the intra- and extra- Schengen air traffic, and by entry restrictions that span beyond the mere suspension of direct flights to and from China. As for the national air traffic, it could have been maintained unrestricted for a longer timeframe: a finding to be read in the context of larger and earlier testing, larger and earlier volume of police checks, better awareness of the public, increased control on the infections potentially coming from abroad.

\subsection{Taiwan} 
Although government actions in Taiwan successfully prevented exponential growth of confirmed cases, the optimised model predicts earlier actions would have damped growth even further as outlined in {Fig.} \ref{fig:optTW}. By allowing movable dates on border control decisions, Taiwan would have benefited from even earlier actions against China (by a few days) and earlier self-health management and home quarantine regulations against selected nations (those shaded in Table 7 of Supplementary Material in February and March, prior to $19^{\mbox{th}}$ March by the maximum 10 days allowed). This allows a later date for the travel restriction on all foreign nationals that was implemented on  $19^{\mbox{th}}$ March, as well as a slight delay in the decrease of international flight volume. The optimizer increases testing volume 2 weeks earlier than reality and suggests community surveillance and mandatory testing for incoming travellers in the first week of the observatory period. Actions regarding mitigating internal spread are also suggested earlier - heightened hygiene practices, self-health management of positive cases' contact and social distance guidelines should have occurred at least 3 weeks earlier (comparison of {Fig.} \ref{fig:optTW} and Table 12 in Supplementary Material). In addition, national flight volume should be kept low to inhibit internal spread. This allows less people to stay at home as illustrated by the reduced online presence. Finally we see people awareness and quarantine control can both be delayed, the latter significantly. This suggests the Taiwan public awareness is sufficient, and enforcement measures like fines and mandatory quarantine may not be necessary.

\subsection{Comparison} 
Care must be taken in comparing the growth and government actions for two different territories. All data is not equal; different testing methods, diagnostic criteria and methods of counting confirmed cases can affect comparisons. There may also be underlying themes behind data / variables, e.g., the availability of personal protection equipment like masks to enable adherence to control measures and keep health workers safe (3 workers from healthcare settings where reported positive by TCDC \cite{TaiwanCDC}, while in Italy more than 28,000 healthcare workers contracted the virus by the end of April \cite{sole24ore, anelli2020italian}), the method and extent of a region's track and trace program in locating high risk individuals, cultural / social response of a population to government actions, healthcare capacity and quality in early isolation and treatment (0.86 vs 3.02 ICU beds per 10,000 population in Italy \cite{angelico2020covid} and Taiwan \cite{TaiwanHandW} respectively). To overcome such inequity of data in this study, Italy and Taiwan were simulated independently. Variables were chosen to be those for which there was data readily available and have a) known or suspected effect on growth curves or b) an ability to highlight any differences between the two territories. Any overarching themes are then considered.
\vspace{0.5em}

\noindent {\bf Testing}: In volume of testing per number of confirmed cases, Taiwan exceeded Italy \cite{owidcoronavirus}. In volume of testing per population, the roles reverse while Taiwan began testing earlier. Regardless, in modelling both regions, the same recommendation resulted from optimization, higher testing volume (and earlier). This agrees with results from a recent model that distinguishes between tested and untested infections cases \cite{giordano2020modelling}, benefits of systematic testing \cite{peto2020covid} as well as statistical observations from South Korea \cite{sKMeasures}: extensive testing with track and trace and social distance restrictions may contain the growth to an extent where a national lockdown can be avoided. 
\vspace{0.5em}

\noindent {\bf Border control:} Either a reduction in international flights is required or a strong border control measure. We see these two measures `meet' in {Fig.} \ref{fig:optIT} for Italy, where the optimizer only allowed 10 day shifts. The border control in optimized Taiwan doesn't drop below $0.7591$ (which is obtainable by 10 day shifts), which is roughly satisfied by mandatory 14 day home quarantine for incoming travellers from infected areas (pending on incoming / outgoing traveller ratios). This allows slightly more international flights ({Fig.} \ref{fig:optTW}). These translate to earlier action against other nations than China for both Italy and Taiwan. 
\vspace{0.5em}

\noindent {\bf People awareness and compliance:} In Taiwan, enforcement of regulations appears not to be a driving factor while people awareness is seemed to be almost ideal. Italy, on the other hand, would have benefited from earlier awareness by the public and earlier implementation of quarantine control. These results seem to highlight the cultural and social differences of the two territories and can be justified empirically using the well-known Hofstede dimensions \cite{hofstede2011dimensionalizing}. Specifically, with a score of 50 in the Power Distance dimension, Northern Italy tends to prefer equality and a decentralisation of power and decision-making. Control and formal supervision is generally disliked among the younger generation. Furthermore, the high score obtained in the Individualism dimension (76 points) accentuates the aversion of the Italian citizen to being supervised and limited in his/her autonomy. Conversely, Taiwan is a hierarchical society in which people accept that hierarchical order and needs no further justifications (the scores in the Individualism and Power Distance dimensions are 17 and 58 points, respectively). 
\vspace{0.5em}

\noindent{\bf National lockdown:} Taiwan never implemented a national lockdown and our results support this decision, even hinting that the observed self-inflicted increase in home activities carried out by the public was not necessary. However maintaining hygiene and social distancing while in public was deemed important and should have happened earlier. Italy, alternatively may have delayed or reduced their nationwide lockdown as a result of increasing testing, border control, quarantine enforcement, public awareness and restrictions localized in the most affected areas.
\vspace{0.5em}

\noindent{\bf Limitations:} It must be noted that neural networks have limitations as they behave as stochastic processes. It should therefore be highlighted that this study further complements existing results by mathematical modelling and statistical analysis. That a data-only driven model arrives at the same results, further fortifies their conclusions. It should also be noted that \emph{hindsight is 20-20}. This study does not serve to criticise governments' management of the pandemic but to assist in preventing a second wave. In particular at the time of most national lockdowns: China was the only example of a nation that had recovered from exponential growth (accomplished by a nationwide lockdown); a large part of the research this study supports was not yet complete; the power of testing and border control illustrated by South Korea and Taiwan respectively had not been fully witnessed; and knowledge of the virus and how it spread via pre-symptomatic individuals or reservoirs was not yet known. However, this exemplifies that as research continues and expands, so will our knowledge and with that more and more tools to overcome this pandemic.

\section{Conclusion} \label{sec:conclusion}

This is the first purely data-driven study where quantitative and qualitative data have been merged and used in a multivariate time series model aimed at predicting the daily increase of the cumulative infected people. We provide an updated and extensive epidemiology overview of both Italy and Taiwan. The purely data-driven approach yielded results that are in line with the main findings of the epidemiological models, proving that the approach has value and that the data alone contains valuable knowledge to inform decision makers. 
    
An in depth discussion of the results obtained has been provided for each of the territories analysed: our main finding being earlier and larger deployment of testing among the population and national entry restrictions allow an easing of lockdown. These are combined with socio-cultural identification of driving factors in each individual territory; stricter enforcement of regulations and information in Italy and relaxation of control measures and self-imposed home confinement in Taiwan. These results give scientific insight to the impact of government actions, fortifying those results previously shown with purely epidemiological models. Hence, this strengthens the foundation for policy makers in constraining growth in current outbreaks and future potential second waves.

This first proof of principle can be built on in several dimensions. One can include more variables, into the time series, for example the availability of ICU beds or personal protection equipment (masks, facial shields). A similar study can be completed for other countries that have experienced a different trend in infections growth rate and that have implemented differently containment measures, e.g., Sweden, United States, Australia, New Zealand, etc. The different models can be analysed in comparison to further increase our understanding of the relation between government actions and social behaviour, and the growth of the epidemics.

This study, along with further developments, in turn can be used to build on existing epidemiology models with the addition of new variables and their fine-tuned effects.


%


\ifCLASSOPTIONcompsoc
  \section*{Acknowledgments}
\else
  \section*{Acknowledgment}
\fi

The authors would like to thank all individuals that have and continue to partake in the global effort of collecting epidemiological data of the SARS-CoV-2 pandemic. In particular the Taiwan CDC, the Italian Department of Civil Protection and the John Hopkins University Center for Systems Science and Engineering for making so much data publicly available. We thank Tom Gheysens for insightful discussions.

AR is funded by SFC Scottish Funding Council. JG is funded by the ERC Consolidator Grant 773202 ERC-2017-COG ‘BabyRhythm’. FFN is funded by the Spanish Ministry of Science under Project ENE2017-88889-C2-1-R. AH is is funded by the Natural Sciences and Engineering Research Council of Canada. Research at Perimeter Institute is supported in part by the Government of Canada through the Department of Innovation, Science and Economic Development Canada and by the Province of Ontario through the Ministry of Colleges and Universities. 

The authors declare no competing interests. Data will be made available via github prior to publication.

\ifCLASSOPTIONcaptionsoff
  \newpage
\fi



%

\bibliography{covidBib}
\bibliographystyle{IEEEtran}



%








\newpage
\clearpage


\section{Supplementary Material: Introduction}\label{sec:introduction}

%
%
%
%
\IEEEPARstart{T}{his} is the supplementary material supporting the paper, \emph{Optimisation of non-pharmaceutical measures in COVID-19 via nerual networks}. We give further details on the data gathering in both description and methods of quantification, which are summarised in the Tables and Figure provided below. 

Sec. \ref{sec:border} summarises the measures taken by Italy and Taiwan in restricting entry and quarantining incoming travellers. We describe in detail the method used to quantify these restrictions as outlined in the Tables \ref{tab:levels}, \ref{tab:Italy_borderControl} and \ref{tab:Taiwan:border}. Other variables that required more details and an outline of actions are described in Sec. \ref{sec:other}. These include enforcement (Taiwan only), testing criteria, government mitigation and population awareness as recapped in Tables \ref{tab:Enforcement_Taiwan} (Taiwan enforcement), \ref{tab:Italy:TestingCriteria} \& \ref{tab:Taiwan:TestingCriteria} (testing criteria), \ref{tab:MitigationActs_Italy} \& \ref{tab:MitigationActs_Taiwan} (government mitigation), and \ref{tab:people awarness} (population awareness).

\section{Supplementary Material: Border Control} \label{sec:border}
\subsection{Level descriptions}
The different levels of action applied to travellers entering the country were equated to a numeric scale. This ranges from 0 to 5 for national citizens and 0 to 7 for non-nationals as depicted in Table \ref{tab:levels}, where 0 represents zero constraints and 7 entry banned. 
 
 Level 1 involves quarantine inspections that targets symptomatic travellers on direct flights from an infected area. Level 2, whilst still targeting symptomatic passengers, opens to those with a travel history to an infected area within 14 days. Level 3 requires health declaration forms from an infected area by all travellers, therefore expanding the target to asymptomatic travellers \cite{TaiwanCDChq}.
 
 Level 4 indicates a 14-day self-health management; in Tawian the TCDC describes this as: hand and respiratory health, record temperature and activities twice daily, avoid public places (wearing a face mask if necessary), and contact local health bureaus if fever or respiratory symptoms develop (inform physician of any history of travel, occupation, contact, and cluster) \cite{TaiwanCDCsh}.
 
 Level 5 requires individuals to a 14-day home quarantine; again for Taiwan the TCDC required travellers to follow several measures \cite{TaiwanCDChq}; mask on arrival, private transportation from airport/point of entry, hand and respiratory health, stay at home, keep one meter away from those sharing the same household and avoid contact with them. Quarantine/isolation hotels can be availed of if one's home/destination does not qualify for home quarantine. Groceries or necessary products are to be obtained via help of household or family members, alternatively one can contact the care and support center for quarantined/isolated individuals. Non-urgent medical care should be postponed and if symptoms develop, individuals are not to seek medical attention by themselves but contact health authorities to arrange their medical care.
 
Level 6 has a restriction on entry visas for specified foreign nationals while level 7 involves a ban on entry for non-residents. 
 
 \begin{table*}[h]
    \centering
    \begin{tabular}{||c | c c|| }
        \hline \hline
        Levels & Description of person applicable & Action  \\ 
        \hline \hline
        0 & Everyone & None  \\ 
        \hline
        1 & Symptomatic from infected area & Quarantine Officer \\
        \hline
        2 & Symptomatic from infected area ($<14$ days) & Quarantine Officer \\
        \hline
        3 & Asymptomatic from infected area ($<14$ days) & Health Card Declaration \\
        \hline
        4 &  & 14 days self-health management \\
        \hline
        5 &  & 14 days quarantine \\
        \hline
        6 &  & Reduced entry (visa restrictions) \\
        \hline
        7 & & Entry prohibited \\
        \hline \hline
    \end{tabular}
    \caption{Levels of action taken by a region to protect its border. Levels 6 and 7 cannot apply to citizens or permit holders (legal residents, work permit holders, etc.) of said area.}
    \label{tab:levels}
\end{table*}

 
 \subsection{Italy}
 On January 31st, the Italian Presidency of the Council of Ministers, the Ministry of Health issued the suspension of direct flights between Italy and China. After this decision, other measures of increasing severity have been taken to control the cross-border traffic. Data regarding these legal acts was collected from the archive of regulations issued by the Ministry of Health \cite{trovanorme_minsalute}. Dates and detail are listed below in Table \ref{tab:Italy_borderControl} along with the suitable quarantine levels as described in Table \ref{tab:levels}.
 

\begin{table*}[h]
    \centering
    \begin{tabular}{||c |p {9 cm}|p {3.5cm}| p {0.9cm}|| }
        \hline \hline
        Date & Description & Travel From & Level  \\ 
        \hline \hline
        20/01/2020 &Temperature measured for travellers coming from China. & China & 2 \\ 
        \hline \rowcolor{Gray}
        31/01/2020 & 
        Direct flights from China ceased.  & China (Italian) \newline China (Chinese) & 5 \newline 7 \\\hline \rowcolor{Gray}
        05/02/2020 & 
        Temperature measured for all incoming travellers.  & Anywhere & 2 \\\hline
        21/02/2020 & 
        Quarantine for travellers (symptomatic or not) coming from anywhere after having stayed in China in the last 14 days.  & Anywhere & N/A \\ \hline \rowcolor{Gray}
        17/03/2020 & Quarantine for travelers (symptomatic or not) coming from anywhere & Anywhere &  5 
        \\\hline
        \hline \hline    
        \end{tabular}
    \caption{Border control measures initiated by Italy. Shaded regions are actions, the dates of which are varied in optimisation.}
    \label{tab:Italy_borderControl}
\end{table*}
 
 
 \subsection{Taiwan}
 Taiwan initiated on-board quarantine inspections on the same day China reported cases of pneumonia of an unknown etiology in the city of Wuhan to the World Health Organisation (WHO), $31^{\mbox{st}}$ December 2019. From there they adapted to the global spread of the virus, with different requirements rolled out for different regions of the world as outlined in Table \ref{tab:Taiwan:border}. As the epidemic in China worsened, Taiwan increased entry requirements for different provinces of China. Initially they focused on symptomatic travellers from Wuhan, escalating to an entry ban for Wuhan residents on $23^{\mbox{th}}$ January, with severe restrictions on all Chinese citizens following on $26^{\mbox{th}}$ January. Early February saw an entry ban for Chinese citizens and foreign nationals with a travel history in China, Macau or Hong Kong, with home quarantine for Taiwan, Macau and Hong Kong residents. By end of February, as the virus spread globally, Taiwan initiated a variation of self-health management and home quarantine for travellers from Thailand, Italy, Iran, Singapore, Japan and South Korea. In March, as the situation in Europe worsened, Taiwan further enhanced its measures by mandatory quarantine for incoming travellers from Schengen, U.K., Ireland, and Dubai. This was quickly followed by several Asian countries, Moldova, U.S., New Zealand, Australia and Canada. On $19^{\mbox{th}}$ March, Taiwan announced banned entry for all  foreign national travellers without documents of granted entry. 
 
 In addition to these actions, in rare cases where it was acknowledged that mandatory quarantine orders had been issued later than ideal, retrospective health monitoring was put in place for recently arrived travellers from specified regions. Other scenarios that required spacial attention to which the Central Epidemic Command Center (CECC) reacted include: the docking of the Princess Diamond where the use of smart technology assisted in identifying and following up with 627,386 individuals \cite{info:doi/10.2196/19540}; special requirements in dealing with Taiwan residents on other cruise ships; evacuating Taiwan residents from Wuhan \cite{LEE2020392}; and an outbreak on a three-ship fleet, Dunmu (敦睦) Fleet.
 
 This is summarized in Table \ref{tab:Taiwan:border}, along with the allocated level as described in Table \ref{tab:levels}. Some information was not included in the analysis, yet still included in Table \ref{tab:Taiwan:border} for completeness. 
 

\begin{table*}[h]
    \centering
    \begin{tabular}{||c |p {9.6 cm}|p {4cm}| p {0.9cm}|| }
        \hline \hline
        Date & Description & Travel From & Level  \\ 
        \hline \hline
        31/12/2019 & Onboard quarantine inspection of direct flights. & Wuhan & 1 \\ 
        \hline
        06/01/2020 & 
        Quarantine inspection of passengers. Passengers report if symptomatic within 10 days of return.  & Wuhan \& nearby & 1 \\\hline
        09/01/2020 & 
        Report if symptomatic within 14 days of return.  & Wuhan \& nearby & 2 \\\hline \rowcolor{Gray}
        23/01/2020 & 
        Entry denied for residents of Wuhan.  & Wuhan (res) & 7 \\\hline \rowcolor{Gray}
        24/01/2020 & 
        Stay home (wear mask if one must leave). \newline 
        Avoid crowded places and public transport (mask if must
        ) \newline
        Novel Coronavirus Health Declaration Card
        & Wuhan (non-res) \newline China  \newline China,Macau,HK & 5 \newline 4  \newline  3\\ \hline
        25/01/2020 &
        Report if symptomatic within 14 days of arrival &
        Everywhere & 2 \\ \hline \rowcolor{Gray}
        26/01/2020 &
        Extensive visa restrictions on Chinese nationals  &
        China (res) & 6 \\ \hline
        01/02/2020 &
        Those returning from China to avoid hospitals and healthcare facilities for 14 days. Healthcare professionals to halt work for 14 days with self-health management  & 
        China & N/A \\\hline
        02/02/2020 &
        Entry denied for residents of Guangdong \newline
        14 day home quarantine: entry via cross-strait airports w/ Guangdong travel history (14 days)  \newline
        14 days of self-health management &
        Guangdong (res) \newline Guangdong (res)  \newline \newline China, Macau,HK & 7 \newline N/A \newline \newline 4 \\ \hline
        03/02/2020 &
        Entry denied for residents of Wenzhou \newline
        Home quarantine: travel history to Wenzhou \newline
        Extraction of Taiwanese nationals from Wuhan by chartered plane: 14 day group quarantine &
        Wenzhou (res) \newline Wenzhou (non-res) \newline Wuhan (Taiwanese) & 7 \newline 5 \newline  N/A \\ \hline
        04/02/2020 &
        Home quarantine: travel history to Zhejiang &
        Zhejiang & 5 \\ \hline
        05/02/2020 &
        Cruise ships docked in China, Macau, HK within 14 days banned from docking \newline
        Cruise ships with suspected cases within 28 days banned from docking &
        China, Macau, HK \newline \newline Everywhere & N/A \newline \newline N/A \\ \hline
        06/02/2020 &
        14 day home quarantine for Taiwan nationals, \newline \newline 14 day self-health management for Taiwan nationals permitted entry \newline
        Chinese nationals forbidden entry &
        China, Macau, HK (Taiwanese) \newline Macau, HK (Taiwanese) \newline China (res) & 5 \newline \newline N/A \newline 7 \\ \hline
        07/02/2020 &
        14 day quarantine for Macau and HK nationals \newline
        Entry forbidden for foreign nationals with travel history within 14 days to China, Macau, HK &
        Macau, HK (res) \newline China, Macau, HK (foreigners) & 5 \newline 7\\ \hline
        08/02/2020 &
        Taiwanese passengers on infected cruise ships allowed to return after 2 negative tests, followed by 14 day self-health management on return &
        Everywhere (Taiwanese) & N/A \\ 
        \hline 
         10/02/2020 &
        14 day quarantine: passengers transiting thro' &
        China, Macau, HK & N/A \\ \hline
        11/02/2020 &
        Health declaration form &
        Everywhere & 3 \\ \hline
        13/02/2020 &
        Foreign nationals prohibited transiting or entry &
        Cruise ships & N/A \\ \hline
        14/02/2020? &
        Passenger Health Declaration and Home Quarantine Information System (Entry Quarantine System) &
        Everywhere & N/A \\ \hline
        15/02/2020 &
        14 day home quarantine: avoid public transport from airport (wear mask if must) &
        Endemic (China)  & N/A \\\hline
        22/02/2020 &
        Taiwanese passengers on infected cruise ships allowed to return after 2 negative tests, followed by 14 day group quarantine on return &
        Everywhere (Taiwanese) (Princess Diamond) & N/A \\ \hline \rowcolor{Gray}
        24/02/2020 &
        14 day self-health management & 
        Thailand, Italy, Iran, Singapore, Japan & 4 \\ \hline \rowcolor{Gray}
        25/02/2020 &
        14 day home quarantine from South Korea (SK) \newline
        14 day self-health management from SK &
        SK (foreign) \newline
        SK (Taiwanese) & 5 \newline 4 \\ \hline \rowcolor{Gray}
        27/02/2020 &
        14 day home quarantine from SK &
        SK (Taiwanese) & 5 \\ \hline
        28/02/2020 &
        14 day home quarantine &
        Italy & 5 \\ \hline \rowcolor{Gray}
        02/03/2020 &
        14 day home quarantine (not including transits)\newline
        14 day home isolation &
        Iran \newline Infected flight
        & 5 \newline N/A \\ \hline \rowcolor{Gray}
        07/03/2020 &
        14 day self-health management &
        France, Germany, Spain & 4 \\ \hline \rowcolor{Gray}
        11/03/2020 &
        14 day self-health management &
        Schengen, Bahrain, Kuwait & 4 \\ \hline \rowcolor{Gray}
        14/03/2020 &
        14 day home quarantine \newline
        14 day self-health management & 
        Schengen, UK, Ireland, Dubai \newline Everywhere 
        & 5 \newline 4 \\ \hline 
        17/03/2020 &
        14 day home quarantine &
        Asia19, Moldova, California, Washington, New York & 5 \\\hline
        18/03/2020 &
        14 day home quarantine &
        New Zealand, Australia, Canada, U.S. & 5 \\\hline \rowcolor{Gray}
        19/03/2020 &
        14 day home quarantine \newline
        Entry banned without entry permit &
        Everywhere \newline Foreign nationals & 5 \newline 6 \\\hline
        24/03/2020 &
        Suspension of transit of airline passengers through Taiwan &
        Everywhere &  N/A \\ \hline 
        01/04/2020 &
        Home quarantined travellers prohibited from traveling to offshore islands by plane or boat. \newline
        Residents of offshore islands urged to undergo home quarantine and related measures on the main island &
        Everywhere & N/A \\ \hline
        03/04/2020 &
        Travellers symptomatic within 14 days to be tested and use designated transport vehicles to a designated location for home quarantine. &
        Everywhere & N/A\\ \hline 
        18/04/2020 &
        Inbound travelers who have visited Europe and the Americas within 14 days should voluntarily present documents for home quarantine requirements before boarding. \newline
        Such travelers should stay at quarantine hotels if they live with increased-risk individuals &
        Everywhere & N/A \\ \hline \hline
    \end{tabular}
\end{table*}

\begin{table*}[h]
    \centering
    \begin{tabular}{||c |p {9.6 cm}|p {4cm}| p {0.9cm}|| }
        \hline \hline
        Date & Description & Travel From & Level  \\  \hline
        19/04/2020 &
        Hospital isolation for positive cases, group quarantine for all others. &
        Members of navy from infected ships & N/A\\ \hline
        21/04/2020 &
        Inbound travelers who have visited Southeast Asia within 14 days should complete the COVID-19 Health Declaration and Home Quarantine Notice and confirm if their residence satisfies home quarantine requirements before boarding. \newline
        Such travelers should stay at quarantine hotels if they live with increased-risk individuals. &
        Everywhere & N/A \\
        \hline \hline
    \end{tabular}
    \caption{Border control measures initiated by Taiwan. Date of action and border control level designation as described by Table \ref{tab:levels}. Quarantine inspection implies fever screening of arriving passengers, screening suspected cases through inquiring about their history of travel, occupation, contact, and cluster and conducting health assessments. In applying levels to population in this study, ``Wuhan'' and ``Wuhan + nearby'' are taken to be the Hubei province. Wenzhou city is taken as the Zhejiang province. N/A notes information that was not included in the analysis but which we list for completeness. HK and SK are shorthand for Hong Kong and South Korea respectively. Asia19 refers to 19 Asian countries:  Bangladesh, Bhutan, Brunei, Cambodia, East Timor, India, Indonesia, Japan, Laos, Malaysia, Maldives, Myanmar, Nepal, North Korea, the Philippines, Singapore, Sri Lanka, Thailand, and Vietnam. Infected navy ships refers to Navy serving on a three-ship fleet, Dunmu (敦睦) Fleet. Shaded regions are actions, the dates of which are varied in optimisation}
    \label{tab:Taiwan:border}
\end{table*}
 
 
  \subsection{Quantifying border control}
 To quantify these actions for both Taiwan and Italy, each territory is allocated a quality of control between 0 and 1 each day, thus creating a time series for both regions as illustrated in Figure. \ref{fig:BorderControl}. To compute this level of control on a given day for either territory, the following procedure was followed using available data from John Hopkins University \cite{covid-dataset}.
 
 The daily cases of every country reported globally have been collected and smoothed by a rolling 3-day average (previous, current, day after). The percentage of daily cases associated with each country is then computed, with respect to the global daily cases, and excluding daily cases in Taiwan and Italy respectively (as it is their protection from incoming cases being gauged). For Taiwan, areas associated with China that were treated differently, i.e., different regulations on different dates, include Hubei, Guangdong, Zhejiang, China (rest), Hong Kong and Macau. When actions were directed by cities, provincial data was used. In addition, some states of the United States (U.S.) were treated differently, namely Washington, New York and California. As actions differed only by a day, the date of regulations affecting these states have been taken to represent all of the U.S.
 
As different levels of quarantine are implemented for residents and non-residents (see Table \ref{tab:Taiwan:border} and \ref{tab:Italy_borderControl} for Taiwan and Italy respectively). Statistics from the Taiwan Tourism Bureau \cite{TaiwanTB} for 2019 have been used to encapsulate the ratio of travellers that are residents and visitors as outbound and inbound respectively. For China, the national statistics were used for each province. Similar information was obtained from the World Tourism Organisation for Italy visitors \cite{doi:10.5555/unwtotfb0380011120142018201910} and returning residents \cite{doi:10.5555/unwtotfb0380250119952018202001} in 2018. This data has been used to obtain a weighted-level time series by dot product of the ratios (residents, non-residents) and the levels associated for that day for each country. For example, with travellers from South Korea to Taiwan
\begin{eqnarray*}
    WL^{(SK)}(n)&=&\sum_{i=1}^{2} L^{(SK)}(n,i)*RT^{(SK)}(i), \\
    \Rightarrow WL^{(SK)}(37)&=&0.4932*1+0.5068*0.7143 \\
    &=& 0.8552,
\end{eqnarray*}
where $L^{(SK)}(n)$ is the level on day n from South Korea (on day 37 or $27^{\rm{th}}$ February we have $L^{(SK)}(37)=(\frac55,\frac57)$ from Table \ref{tab:Taiwan:border}) and $RT^{(SK)}$ is the ratio of travellers (outbound, inbound) with $RT^{(SK)}=(0.4932,0.5068)$ from \cite{TaiwanTB}.

The time series quantifying the border control for say Taiwan entry from each country is calculated by multiplying the percentage of daily cases for a given country on a given day by the weighted-level calculated for that country on that day. Again, for the above example of South Korea on day 37,
\begin{eqnarray*}
    TS^{(SK)}(n)&=&WL^{(SK)}(n)*D^{(SK)}(n),\\
    \Rightarrow TS^{(SK)}(37)&=& 0.8552*0.4092=0.3499,
\end{eqnarray*}
where $D^{(SK)}(n)$ is the smoothed daily global percentage of cases associated with South Korea on day $n$ (with Taiwan cases removed) obtained using John Hopkins data \cite{covid-dataset}. By summing all the time series for each country, the daily total quarantine incoming measure for Taiwan (or Italy) on each day is retrieved. This summing of reactions is illustrated by a stacked bar graph in Figures \ref{fig:sub1} and \ref{fig:sub2} for Italy and Taiwan respectively.

\begin{center}
    \begin{figure*}[h]
        \centering
        \begin{subfigure}{0.99\textwidth}
  \centering
  \includegraphics[width=0.65\textwidth]{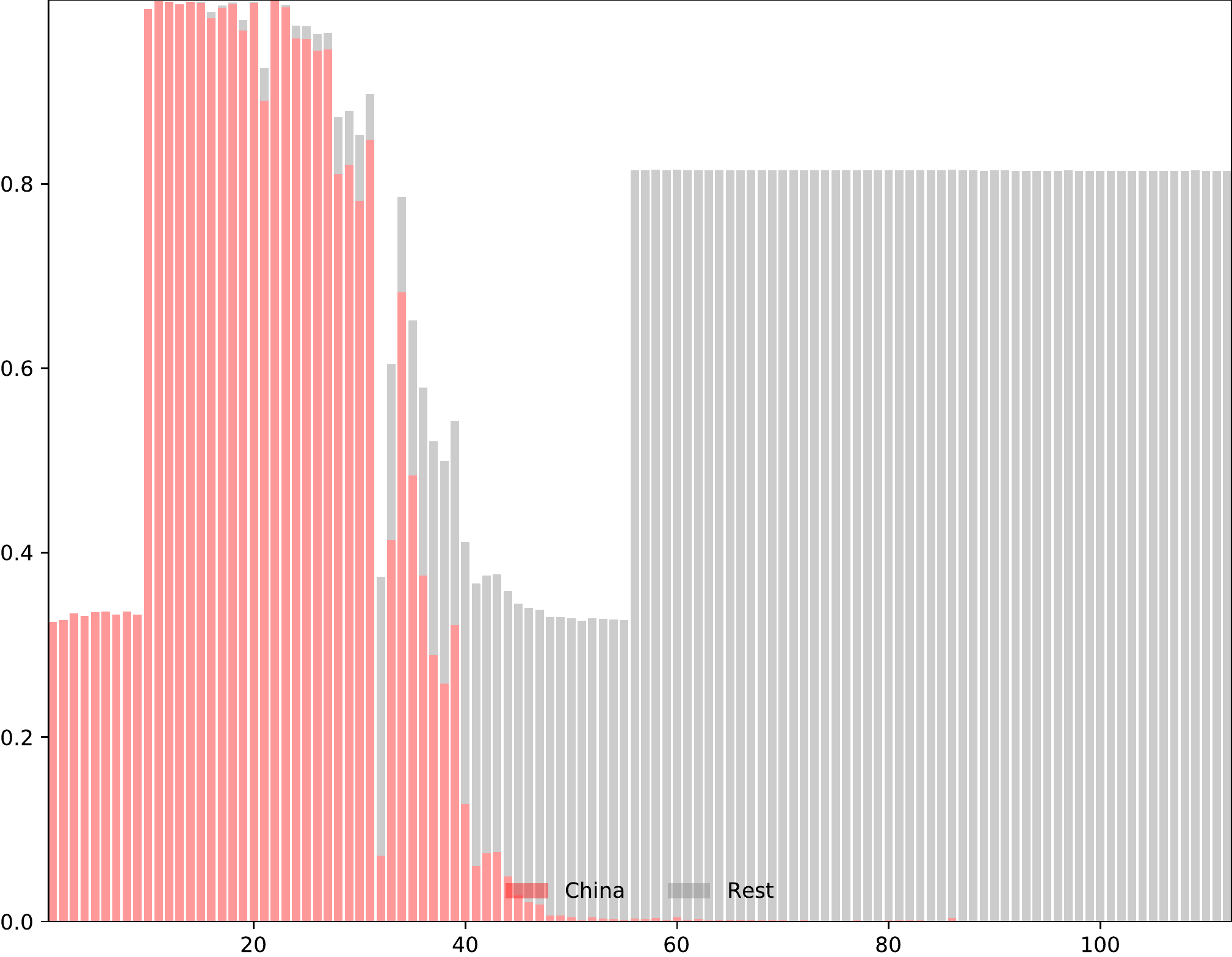}
  \caption{Italy}
  \label{fig:sub1}
\end{subfigure}
\begin{subfigure}{0.99\textwidth}
  \centering
  \includegraphics[width=0.65\textwidth]{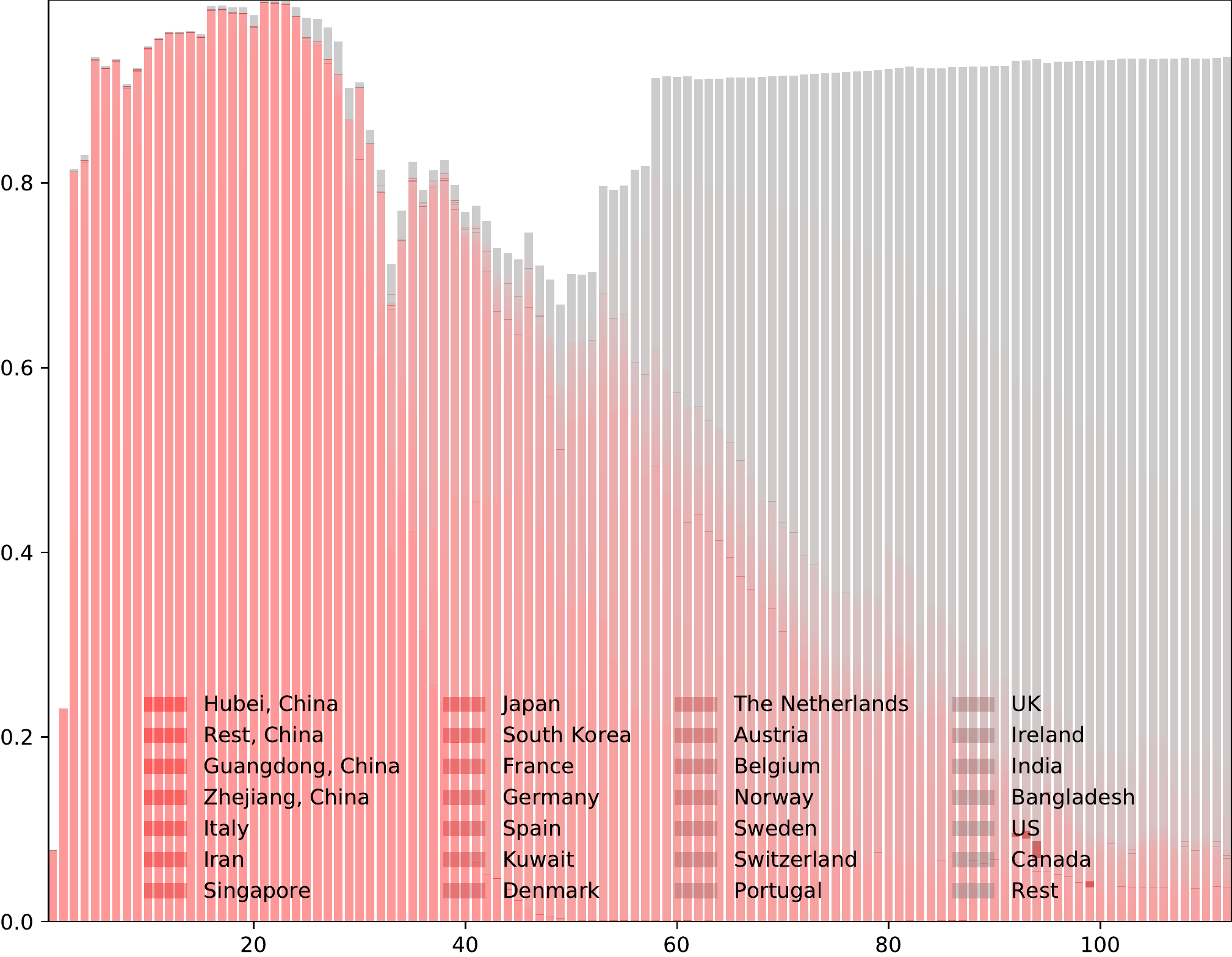}
  \caption{Taiwan}
  \label{fig:sub2}
\end{subfigure}
        \caption{Formation of border control measure for (a) Italy and (b) Taiwan. For each territory, their reaction to different nations and areas are quantified and summed. The x-axis represents time in days starting with $22^{\mbox{nd}}$ January as day 1. The y-axis is the sum of the border control measures, the total ranging from 0 (no measures) to 1 (mandatory quarantine for national residents and entry denied for all others). For visual ease, only the impact of 20 countries are individually shown for Taiwan, in reality, 62 nations were individually calculated.}
        \label{fig:BorderControl}
    \end{figure*}{}
\end{center}

\vspace{-0.7cm}
\section{Supplementary Material: Further government actions and population response} \label{sec:other}

\subsection{Taiwan enforcement}

To encourage compliance with quarantine and isolation orders in Taiwan, fines were introduced early of NT\$150,000 and NT\$300,000 respectively (USD\$5,000 and USD\$10,000) and raised to NT\$1 million (USD\$33,000) on $27^{\mbox{th}}$ March and mandatory group quarantine at a designated centre. Health Declaration Cards were required by travellers from high-risk areas from $24^{\mbox{th}}$ January with smart technologies announced on $28^{\mbox{th}}$; this evolved into the Entry Quarantine System announced on $14^{\mbox{th}}$ February - an online system where travellers could obtain a mobile health declaration pass on their phones. This hastened the immigration process, allowed the Taiwan healthcare system immediate access to travel history of `perspective' future patients and further enabled mobile tracking of quarantined individuals. Inaccurate information earned travellers a NT\$150,000 fine (USD\$5,000). By $3^{\mbox{rd}}$ April, the LINE Bot system was announced to track quarantined individuals, however the newer system also allowed voluntary health reporting and updated information on prevention measures. These were reported to the public \cite{TaiwanCDC} (ongoing) and previously noted in \cite{10.1001/jama.2020.3151} (until $24^{\rm{th}}$ February 2020). Table \ref{tab:Enforcement_Taiwan} lists and quantifies these consequences as a step function.

\begin{table*}[h]
    \centering
    \begin{tabular}{||c|p {14cm}|c||}
        \hline \hline
        Date & Regulation Description & Level \\ \hline \hline
        28/01/2020 &
        Smart technology utilised to provide assistance in prevention. &
        1 \\ \hline
        11/02/2020 &
        Inaccurate information from travellers (refusal, evasion, interference) results in a fine up to NT150,000. &
        2 \\ \hline
        12/02/2020 &
        Violators of home isolation fined up to NT300,000. Violators of home quarantine fined up to NT150,000. &
        3 \\ \hline
        27/03/2020 &
        Fines increased up to NT1 million for violation of quarantine or isolation measures and put into group quarantine. &
        4 \\ \hline
        03/04/2020 &
        LINE Bot system: Disease Containment Expert introduced to track people in home quarantine &
        5 \\ \hline
        05/04/2020 &
        Fine up to NT15,000 for not wearing mask on public transport. &
        6 \\
        \hline \hline
    \end{tabular}
    \caption{Enforcement actions taken in Taiwan.}
    \label{tab:Enforcement_Taiwan}
\end{table*}


\subsection{Testing criteria}

The progression of testing in who qualified for tests for both Italy and Taiwan are summarised in Tables \ref{tab:Italy:TestingCriteria} and \ref{tab:Taiwan:TestingCriteria} respectively below.


\begin{table*}[h]
    \centering
    \begin{tabular}{||c |p {14 cm}| c|| }
        \hline \hline
        Date & Description & Level  \\ 
        \hline \hline
        22/01/2020 & Testing symptomatic travellers coming from China & 1 \\ \hline
        27/01/2020 & Testing symptomatic individuals that had contact with a confirmed patient in the last 14 days & 2 \\\hline
        22/02/2020 & Testing individuals showing symptoms of either \textit{influenza-like-illness} (ILI) or \textit{Severe Acute Respiratory Infection} (SARI), and that has been in contact with a suspect case, or with history of travel in China & 3\\\hline
        09/03/2020 & Testing anyone showing ILI or SARI symptoms and that requires hospitalization & 4\\
        \hline \hline    
        \end{tabular}
    \caption{Decision criteria employed in Italy for diagnostic tests}
    \label{tab:Italy:TestingCriteria}
\end{table*}


\begin{table*}[h]
    \centering
    \begin{tabular}{||c |p {14 cm}| c|| }
        \hline \hline
        Date & Description & Level  \\ 
        \hline \hline
        06/01/2020 &
        Healthcare facilities to reinforce case reporting of severe cases of pneumonia among travellers from Wuhan. &
        1 \\ \hline
        24/01/2020 &
        Travellers from Wuhan with fever or acute respiratory tract infection &
        2 \\ \hline
        12/02/2020 &
        Mandatory testing for severe complicated influenza cases, cases of community-acquired upper respiratory infections under surveillance and clusters of upper respiratory infections who were reported on and after January 31 and whose specimens tested negative &
        3 \\ \hline
        16/02/2020 &
        Community surveillance expanded to include \newline
        1) Any individuals with foreign travel history in the past 14 days or any individuals who have had contact with symptomatic foreign travelers with a fever or respiratory symptoms and highly suspected of having the said COVID-19 symptoms caused in the past 14 days. \newline
        2) Clusters of cases of fever/respiratory symptoms. \newline
        3) Pneumonia cases whose symptoms haven’t improved after three days of antibiotic therapy for unknown cause or clusters of pneumonia cases or healthcare workers having pneumonia. &
        4 \\ \hline
        18/03/2020 &
        Retrospective testing for symptomatic people entering from Europe, Egypt, Turkey, Dubai &
        5 \\ \hline
        21/03/2020 &
        Retrospective testing for symptomatic people entering from US, East Asia &
        6 \\ \hline
        07/04/2020 &
        Mandatory testing for inbound travellers with fever or symptoms in the past 14 days &
        7 \\
        \hline \hline    
        \end{tabular}
    \caption{Progression of surveillance in testing the population of Taiwan}
    \label{tab:Taiwan:TestingCriteria}
\end{table*}

\subsection{Government mitigation}
Actions taken by both governments to implement social distancing and health promotion are listed and quantified for both Italy and Taiwan in Tables \ref{tab:MitigationActs_Italy} and \ref{tab:MitigationActs_Taiwan} respectively.


\begin{table*}[h]
\centering
\begin{tabular}{|| c| p {14 cm}| c ||}
\hline \hline
Date                           & Description      & Level                         \\ \hline \hline
23/02/2020                     & Creation of red zones in Northern Italy            & 1      \\ \hline
25/02/2020                     & Additional limitations in the red zones: suspension of sport events, meetings, school trips          & 2  \\ \hline
1/03/2020                      & Closure of schools, suspension of events of all kinds in the red zones & 3\\ \hline
4/03/2020                      & Closure of schools and universities in the whole country & 4 \\  \hline
8/03/2020                      & "Stay home" Decree Law  & 5 \\ \hline            
11/03/2020                     &  Closure of restaurants, bars and retail stores   & 6       \\\hline
20/03/2020                     & Closure of parks, ban on outdoor sport activities  & 7    \\\hline
22/03/2020                     & Closure of non-essential manufacturing businesses  & 8 \\ \hline \hline
\end{tabular}
\caption{Chronology of legal acts taken in Italy to contain the spread of the epidemics \cite{gazzettaufficiale} }
\label{tab:MitigationActs_Italy}
\end{table*}


\begin{table*}[h]
    \centering
    \begin{tabular}{||c|p {14cm}|c||}
        \hline \hline
        Date & Mitigation Description & Level \\ \hline \hline
        06/01/2020 &
        Healthcare facilities: Workers strictly adhere to standard precautions for preventing nosocomial infection, wearing N95 respirators while performing invasive medical procedures. Heightened vigilance for suspected cases, and thoroughly implement the inquiry of patients for history of travel, occupation, contact, and cluster (TOCC). &
        1 \\ \hline
        20/01/2020 & 
        School holidays begin (later extended 2 weeks). &
        2 \\ \hline
        24/01/2020 &
        Heightened hygiene practices: Healthy public to take temperature, wash hands thoroughly with soap, avoid touching eyes, nose and mouth with hands and avoid crowded public places. If symptomatic, rest at home and stay clear of crowded public places (wear surgical mask if must). &
        3 \\ \hline
        25/01/2020 &
        14 days of self-health management (since contact): Those with possible contact with a known positive case, if symptomatic, report to hotline. &
        4 \\ \hline
        29/01/2020 &
        The CECC issued relevant guidance for public transportation, public gatherings, educational institutions and groups and provided advice on prevention measures to reduce the risk of infection. \cite{cdcPrecautions} &
        5 \\ \hline
        12/02/2020 &
        CECC reiterates close contacts are required to abide by home isolation or quarantine as instructed by the government. &
        6 \\ \hline
        19/02/2020 &
        Public to regularly wash their hands with soap, ensure proper ventilation in their homes, and regularly clean their homes’ interior and exterior. Upcoming school reopenings: CECC reminds schools and public kindergartens to institute appropriate cleaning and disinfection procedures for their classrooms and school buses. Public transportation operators to step up disinfection measures. &
        7 \\ \hline
        23/02/2020 &
        Local environmental protection departments tasked with disinfecting public spaces surrounding schools and public kindergartens and spaces within them which were open to the general public during the winter vacation, such as large hallways, wash basins, restrooms, and playgrounds. (by 23/02). &
        8 \\ \hline
        25/02/2020 &
        Schools reopen (conditional: more than one case in a school results in a 2 week closure). &
        7 \\ \hline
        05/03/2020 &
        CECC issued Guidelines for Large-Scale Public Gatherings in the Wake of the COVID-19 Outbreak (see attachment) to give guidance for organizing public gatherings \cite{cdcGatherings}. &
        8 \\ 
        \hline 
        18/03/2020 &
        Retrospective health monitoring on individuals entering Taiwan from Europe, Egypt, Turkey and Dubai. &
        9 \\\hline
        21/03/2020 &
        Retrospective health monitoring symptomatic for people entering from US, East Asia. &
        10 \\ \hline
        25/03/2020 &
        Indoor events with 100+ people and outdoor gatherings 500+ people to be suspended. Organizers of these gatherings can conduct risk assessments &
        11 \\ \hline
        01/04/2020 &
        Phase 1 social distancing implemented: outdoor 1m+, indoor 1.5m+ separation, wear masks otherwise. &
        12 \\ \hline
        05/04/2020 &
        Public transport riders urged to wear masks at all times and undergo temperature checks before entering bus or MRT station (NT15,000 fine) &
        13 \\ \hline
        06/04/2020 &
        14 self-health management for those who attended temple festival, if unwell, call hotline &
        14\\ \hline
        09/04/2020 &
        Host and hostess clubs and ballrooms suspend operations. &
        15 \\ \hline
        10/04/2020 &
        Crowd control measures imposed at public places where large crowds can gather &
        16 \\ \hline
        14/04/2020 &
        Home quarantine/isolation for incoming from Europe and Americas with quarantine hotels for those who live with increased risk individuals. &
        17 \\
        \hline \hline
    \end{tabular}
    \caption{Mitigation actions taken in Taiwan. Increased risk individuals refers to people over 64 years old, children under 7 years old, persons with chronic disease or persons who don’t have a separate room (including a separate bathroom)}
    \label{tab:MitigationActs_Taiwan}
\end{table*}


\subsection{Population awareness}
The list of keywords used for quantifying public awareness via Google Trends are listed in Table \ref{tab:people awarness}
\begin{table*}[h]
\centering
\begin{tabular}{|| p{8 cm} |p{8 cm}||}
\hline \hline
Italy & Taiwan     \\ \hline \hline
  italia coronavirus, notizie coronavirus, news coronavirus, il coronavirus, coronavirus, covid 19, covid italia, coronavirus covid 19, autocertificazione covid, covid, la quarantena, quarantena coronavirus, quarantena italia, quarantena fine, quarantena, wuhan coronavirus, cina, wuhan cina, virus wuhan, wuhan, virus corona, virus italia, corona virus italia, cina virus, virus &       
  coronavirus taiwan, taiwan, coronavirus update, coronavirus cases, coronavirus, covid 19, covid taiwan, covid-19, covid 19 taiwan, covid,quarantine 中文, quarantine 意思, self quarantine,隔離 英文 quarantine, quarantine, wuhan virus, wuhan coronavirus, wuhan pneumonia, wuhan, virus corona,corona, taiwan virus, virus   \\ \hline \hline
\end{tabular}
\caption{People awareness keywords list}
\label{tab:people awarness}
\end{table*}

\end{CJK}
\end{document}